\documentclass[aps,amsmath,amssymb,showpacs,showkeys]{revtex4-2}
\usepackage[dvips]{graphicx}
\usepackage{palatino}
\usepackage{braket}
\usepackage{xcolor}
\usepackage{orcidlink}
\usepackage{hyperref}
\usepackage{color,soul}
\hypersetup{
  colorlinks=true,
  urlcolor=blue,
  linkcolor=red,
  citecolor=blue
}
\begin{document}
\title{Dynamical system analysis of LRS-BI Universe with $\boldsymbol{f(Q)}$ 
gravity theory}

\author{Pranjal Sarmah \orcidlink{0000-0002-0008-7228}}
\email[E-mail:]{p.sarmah97@gmail.com}

\author{Umananda Dev Goswami \orcidlink{0000-0003-0012-7549}}
\email[E-mail:]{umananda2@gmail.com}

\affiliation{Department of Physics, Dibrugarh University, Dibrugarh 786004, 
Assam, India}

\begin{abstract}
Considering the Universe as a dynamic system, the understanding of its 
evolution is an interesting aspect of study in cosmology. Here, we investigate 
the anisotropic locally rotationally symmetric (LRS) Bianchi type-I (LRS-BI) 
spacetime under the $f(Q)$ gravity of symmetric teleparallel theory equivalent 
to the GR (STEGR) as a dynamical system and try to understand the role of 
anisotropy in the evolution of its various phases. For this work, we consider 
two models, viz., $f(Q) = -\,(Q+ 2\Lambda)$ and $f(Q) = -\, \beta Q^{n}$ to 
study the critical points and stability of the LRS-BI Universe. In both the 
models, it is found that the various phases like radiation-dominated, 
matter-dominated, and dark energy-dominated phases are heteroclinically 
connected, and there is some role of anisotropy in the evolution of these 
phases of the Universe. However, for both the models, there are some 
unphysical solutions too depending upon the values of model parameters 
$\beta$ and $n$ along with the anisotropic parameter $\alpha$.
 
\end{abstract}


\maketitle
\section{Introduction}
The two fundamental tenets of conventional cosmology are the isotropy and 
homogeneity in the Universe, and the Friedmann-Lema\^itre-Robertson-Walker 
(FLRW) metric in the framework of general relativity (GR) along with the 
standard $\Lambda$CDM model successfully explains the majority of its 
aspects. However, many trustworthy observational data sources, including WMAP 
\cite{wmap,wmap1,wmap2}, SDSS \cite{Bessett_2009,SDSS_2005,Tully_2023,Eisenstein_2005}, and Planck
\cite{Planck_2018} have shown some deviations from the principles of 
conventional or standard cosmology. These suggest that the Universe may be 
anisotropic in some ways. Additionally, some investigations \cite{Gum_2007, Ackerman_2007} have 
pointed to a large-scale planar symmetric geometry of the Universe. For such 
geometry of the Universe, the eccentricity of the order $10^{-2}$ can 
match the quadrupole amplitude to the observational evidence without 
changing the higher order multipole of the temperature anisotropy in 
the angular power spectrum of CMB \cite{Tedesco_2006}. The symmetry axis in 
the vast scale geometry of the Universe is also indicated by the polarization 
analysis of electromagnetic radiation that is traveling across great 
cosmological distances \cite{akarsu_2010}.

To address the observed small-scale symmetric properties of the Universe, the 
locally rotationally symmetric (LRS) Bianchi type-I (BI) model of the Universe 
is one of the simplest and most straightforward ways. This model has planar 
symmetry in the spatial section along with a symmetry axis in its metric. 
The metric reflects an ellipsoidal expansion of the Universe and 
supports the CMB data, consisting of two longitudinal directions and one 
transverse direction \cite{Sarmah_2023}. As a result, the LRS-BI metric is a 
good choice for exploring the Universe's possible anisotropy. The Bianchi 
cosmology has recently gained popularity in cosmology research because of 
the technological advancements and mounting evidence of anisotropy from 
observational data. Different features of Bianchi models have received 
extensive studies in the Refs.~\cite{Jaffe, Jaffe1, Jaffe2, Tedesco_2004,
Tedesco_2006, Sarmah_2023, Sarmah_2022,espo,A1,A2,A3,A4,A5}.

Another issue of standard cosmology is the inevitable assumption of some 
exotic energy component, known as dark energy (DE) with the properties of 
negative pressure or negative equation of state (EoS) to explain the 
accelerated expansion of the Universe, which was confirmed in 1988 from the
Type Ia supernovae observations \cite{Riess, Perlmutter_1999}. 
Unfortunately, to date, no observational evidence about the existence of DE 
has been discovered. Moreover, the Hubble tension \cite{Planck_2018, 
Nedelco_2021,Valentino_2021}, the $\sigma_8$ tension \cite{Planck_2018,Poulin_2023}, the coincidence 
problem \cite{Velten_2014, Zheng_2023}, and other theoretical \cite{Helbig_2020,Rajantie_2012,Hawking_1966} and 
observational \cite{Buchert_2016,dark_matter} cosmology-related concerns are not clearly understood within 
the context of GR. Modified theories of gravity (MTGs), based on 
modifications of Einstein-Hilbert action of GR, have recently gained 
popularity among researchers as ways to get rid of concepts like DE. Of all 
those MTGs, the $f(R)$ theory \cite{Li_2007,Gogoi_2022,Stachowski_2017} is the most straightforward and 
effective one \cite{Gogoi_2022,Stachowski_2017}. The cosmological studies of various Bianchi models 
along with different MTGs like $f(R)$, $f(R,T)$, etc.\ can be found in 
Refs.~\cite{Liu_2017,Sahoo_2017, mishra_2019}. Other alternative theories of 
gravity (ATGs), such as the teleparallel theory equivalent to GR (TEGR) 
\cite{Bahamonde_2023,Capozziello_2011} and the symmetric teleparallel theory equivalent to GR (STEGR) 
\cite{Paliathanasis_2024, Lu_2019}, have been developed based on an alternative geometrical 
formulation other than the geometrical basis of GR. These ATGs take into 
account flat space and affine connections while also changing the geometry of 
spacetime without affecting its Lagrangian. The non-vanishing torsion is the 
essential foundation of the TEGR theory, while it is the non-metricity that
forms the base of STEGR theory \cite{Sarmah_2023,Jimenez_2018}.

The $f(Q)$ gravity theory is an extended version of the STEGR theory, where 
$Q$ is the non-metricity scalar term. Recent cosmological studies have 
increased interest in the $f(Q)$ idea among researchers. Various aspects 
of cosmological and astrophysical systems have been studied within the 
framework of $f(Q)$ theory and some of them are found in Refs.~\cite{accfQ1,
solanki_2022,fQfT,fQfT1,fQfT2,fQfT3,deepjc,zhao,gde,lin,cosmography,signa,redshift,
perturb,dynamical1,prd/AD,Q1,Q2,Q3,Q4,Q5,Q6,Koussour_2022, Koussour_2024, Solanke_2023, Shekh_2021, Shekh_2023, Shekh_2024,Shabani_2024,ade}. However, the majority of 
works have been done in the isotropic space with the coincident gauge, while 
relatively little research has been done on the anisotropic cosmology in the 
$f(Q)$ theory. To this end in particular, using the LRS-BI metric, different 
cosmological profiles, including the energy density, equation of state, and 
skewness parameters, have been investigated within the $f(Q)$ gravity theory 
in Refs.~\cite{Sarmah_2023, deepjc}. Another study on the anisotropic Universe 
under $f(Q)$ gravity has been made in Ref.~\cite{Devi_2022}, taking into 
account the Hybrid Expansion Law (HEL) for the average scale factors. 
Therefore, it will be interesting to study the anisotropic characteristics of 
the Universe and comprehend the dynamical evolution of the system in terms of 
$f(Q)$ theory. Through a critical point analysis of the LRS-BI Universe in the 
$f(Q)$ theory of gravity, we wish to investigate the dynamical development of 
the system and assess its stability.

Recasting the cosmological equations into a dynamical system is a potent and 
beautiful method to examine the dynamics of the Universe \cite{Agostino_2018}. 
Numerous cosmological models, including the canonical \cite{Heard_2002, 
Fang_2014} and noncanonical scalar-field models \cite{Fang_2016,Copeland_2005, 
Copeland_2010}, the scalar-tensor theories \cite{Agarwal_2008,Huang_2015}, the 
$f(R)$ gravity theory \cite{Amendola_2007,Carloni_2015,Alho_2016}, and others 
\cite{Odintsov_2018, Odintsov_2017,Li_2018, Hrycyna_2013} have been applied to
study the stability of systems using the dynamical system approach. A recent 
review of the dynamical system approach can be found in 
Ref.~\cite{Bahmonde_2018}. A few works have also been done to analyse the 
cosmological dynamical system in $f(Q)$ gravity theory \cite{De_2023, Lu_2019}.
Nevertheless, similar analyses with the Bianchi metric are very limited and 
hence we intend to study it with the LRS-BI metric to extract some information 
about the anisotropic Universe in $f(Q)$ gravity theory.

The current article is organized as follows. Starting from this introduction  
part, we discuss the basic mathematical formalism of $f(Q)$ gravity in 
Section \ref{sec2}. In Section \ref{sec3}, the field equations for the LRS-BI 
model in $f(Q)$ gravity have been derived. The dynamical system analyses for 
two different $f(Q)$ models have been performed in Section \ref{sec4}. In
Section \ref{aniso} the dynamics of anisotropy have been studied. Finally, the 
article has been summarised with conclusions in Section \ref{con}. 

\section{Basic Equations of $\boldsymbol{f(Q)}$ Gravity Theory}
\label{sec2}
As mentioned in the previous section, the symmetric teleparallelism and 
non-metricity conditions are the basic principles on which the $f(Q)$ gravity 
theory is based. Mathematically, these two principles can be expressed 
respectively as $R^\rho{}_{\sigma\mu\nu} = 0$ and 
$Q_{\lambda\mu\nu} := \nabla_\lambda g_{\mu\nu} \neq 0$. In this theory, the 
associated affine connection can be written as  
\begin{equation} \label{connc}
\Gamma^\lambda{}_{\mu\nu} = \mathring{\Gamma}^\lambda{}_{\mu\nu} +L^\lambda{}_{\mu\nu}, 
\end{equation}
where the first term on the right-hand side is the usual Levi-Civita 
connection and the second term is known as the disformation tensor. This 
disformation tensor can be expressed as  
\begin{equation}
L^\lambda{}_{\mu\nu} = \frac{1}{2} (Q^\lambda{}_{\mu\nu} - Q_\mu{}^\lambda{}_\nu - Q_\nu{}^\lambda{}_\mu) \,.
\end{equation}
Moreover, there is another tensor known as the superpotential tensor, which is
defined by
\begin{equation} \label{P}
P^\lambda{}_{\mu\nu} := \frac{1}{4} \Big( -2 L^\lambda{}_{\mu\nu} + Q^\lambda g_{\mu\nu} - \tilde{Q}^\lambda g_{\mu\nu} -\frac{1}{2} \delta^\lambda_\mu Q_{\nu} - \frac{1}{2} \delta^\lambda_\nu Q_{\mu} \Big),
\end{equation}
Using this superpotential tensor the non-metricity scalar can be obtained as 
\begin{equation} \label{Q}
Q = -\,Q_{\lambda\mu\nu}P^{\lambda\mu\nu} 
\end{equation}
and the action of $f(Q)$ gravity takes the form:
\begin{equation}\label{action}
S = \int \Big[\frac{1}{2\,\kappa}\,f(Q) + \mathcal{L}_m \Big] \sqrt{-g}\,d^4 x,
\end{equation}
where $\kappa = 8\pi G$, $G$ being the Newtonian gravitational constant, 
$g$ is the determinant of the metric $g_{\mu \nu}$ and $\mathcal{L}_m$ 
is the matter part of the Lagrangian. We can obtain the field equations of 
$f(Q)$ gravity theory by varying the action \eqref{action} with respect to 
$g_{\mu \nu}$ as given by 
\begin{equation} \label{FE1}
\frac{2}{\sqrt{-g}} \nabla_\lambda (\sqrt{-g}\,f_Q\,P^\lambda{}_{\mu\nu}) +\frac{1}{2}\,f(Q)\, g_{\mu\nu} + f_Q(P_{\nu\rho\sigma}\, Q_\mu{}^{\rho\sigma} -2P_{\rho\sigma\mu}\,Q^{\rho\sigma}{}_\nu) = -\,\kappa\, T_{\mu\nu},
\end{equation}
where $f_Q$ specifies the derivative of $f(Q)$ with respect $Q$ and 
$T_{\mu\nu}$ be the energy-momentum tensor obtained from the matter 
Lagrangian.  

Further, by using the connection coefficient (\ref{connc}), we can obtain the 
following relation for the curvature tensors akin to the affine connection and 
Levi-Civita connection: 
\begin{equation}
R^\rho{}_{\sigma\mu\nu} = \mathring{R}^\rho{}_{\sigma\mu\nu} + \mathring{\nabla}_\mu L^\rho{}_{\nu\sigma} - \mathring{\nabla}_\nu L^\rho{}_{\mu\sigma} + L^\rho{}_{\mu\lambda}L^\lambda{}_{\nu\sigma} - L^\rho{}_{\nu\lambda} L^\lambda{}_{\mu\sigma}
\end{equation}
and thus
\begin{align*}
R_{\sigma\nu} &= \mathring{R}_{\sigma\nu} + \frac{1}{2} \mathring{\nabla}_\nu  Q_\sigma + \mathring{\nabla}_\rho L^\rho{}_{\nu\sigma} -\frac{1}{2} Q_\lambda L^\lambda{}_{\nu\sigma} - L^\rho{}_{\nu\lambda}L^\lambda{}_{\rho\sigma}, \nonumber \\[5pt]
R &= \mathring{R} + \mathring{\nabla}_\lambda Q^\lambda - \mathring{\nabla}_\lambda \tilde{Q}^\lambda -\frac{1}{4}Q_\lambda Q^\lambda +\frac{1}{2} Q_\lambda \tilde{Q}^\lambda - L_{\rho\nu\lambda}L^{\lambda\rho\nu} \,.
\end{align*}
Now, with the implementation of the symmetric teleparallelism condition, the 
field equations in (\ref{FE1}) can be reduced to the form:
\begin{equation} \label{FE2}
f_Q \mathring{G}_{\mu\nu} + \frac{1}{2}\,g_{\mu\nu}(f(Q)-Qf_Q) + 2f_{QQ} \mathring{\nabla}_\lambda Q P^\lambda{}_{\mu\nu} = -\,\kappa\, T_{\mu\nu},
\end{equation}
where $\mathring{G}_{\mu\nu}$ is the Einstein tensor. With these field 
equations, we can proceed to study the LRS-BI Universe.

\section{Field Equations in the LRS-BI model}\label{sec3}
Here we consider the LRS-BI metric in Cartesian coordinates with geometric 
units as given by 
\begin{equation}\label{metric}
ds^2 = -\,dt^2 + a_1 dx^2 + a_2 (dy^2 + dz^2),
\end{equation}
where $a_1$ and $a_2$ are the directional scale factors, specifying 
respectively a transverse direction along $x$-axis and two equivalent 
longitudinal directions along $y$ and $z$-axis. Further, this metric is a 
coincident gauge. The directional Hubble parameters describe by this metric 
are $H_x=\dot{a_{1}}/{a_{1}}$, $H_y=\dot{a_{2}}/{a_{2}}$. So, the average 
Hubble parameter for the LRS-BI Universe can be written as
\begin{align}
(H_x+2H_y)/3=H=\dot{a}/{a},
\end{align} 
where $a= (a_1 a_2^2)^{1/3}$ is the average scale factor for the considered 
anisotropic Universe. Moreover, the shear scalar $\sigma$ appears due to 
anisotropy of the space can be defined by 
\begin{equation}\label{sigma}
\sigma^2=\frac{1}{3}(H_x-H_y)^2=3(H-H_y)^2.   
\end{equation}
The non-metricity scalar $Q$ for the LRS-BI metric can be calculated as 
\begin{equation}\label{Q}
Q=2H_y^2+4H_xH_y.
\end{equation}
$Q$ can also be expressed in terms of shear scalar as
\begin{equation}\label{Qnew}
Q=6H^2-2\sigma^2.
\end{equation}
Using the metric \eqref{metric} along with the definitions of Hubble 
parameters in the field equations \eqref{FE2}, the following set of equations 
of motion can be obtained:
\begin{equation}\label{eom1}
\kappa \rho = \frac{f(Q)}{2} - 2f_Q \Big[2 H_xH_y + H_y^2 \Big], 
\end{equation}
\begin{equation} \label{eom2}
\kappa p_x = -\,\frac{f(Q)}{2} +2 f_Q \Big[ 3HH_y + \dot{H}_y \Big] +2H_y \dot{Q} f_{QQ}, 
\end{equation}
\begin{equation} \label{eom3}
\kappa p_y = -\,\frac{f(Q)}{2} + f_Q \Big[ 3 H(3H-H_y) + \dot{H}_x+\dot{H}_y \Big] + \Big( H_x +H_y \Big) \dot{Q} f_{QQ}, 
\end{equation}
where $\rho$ is the  matter density, and $p_x$ and $p_y$ are the pressures of 
the field along $x$ and $y$ directions respectively. 

One can see that above three field equations (\ref{eom1}, \ref{eom2}, 
\ref{eom3}) contain five unknowns $H_x,H_y,\rho, p_x,p_y$ and the model 
$f(Q)$. Therefore, to solve this system of equations we need some additional 
assumptions. Here we have taken two assumptions out of which the first one is 
in the form of the barotropic EoS, $p=\omega \rho$ and the second one is the
proportionality of the shear scalar $\sigma$ and expansion scalar $\theta$, 
i.e.~$\sigma^2\propto\theta^2$. The latter condition is the simplest 
physically relatable condition, which is well-known in the literature for the
anisotropy-based cosmology \cite{B1,B2,B3,B4,B5,B6,B7,jcap/AD}. Under this 
proportionality condition, we have the directional Hubble parameters relation
\begin{equation}\label{H_directional}
    H_x=\alpha H_y,
\end{equation}
where $\alpha$ is a constant and the average Hubble parameter becomes 
\begin{align}
H=\frac{(2+\alpha)}{3}\,H_y.
\end{align}
Consequently equations \eqref{Q}, \eqref{eom1}, \eqref{eom2} and \eqref{eom3} 
take the forms as follows \cite{Sarmah_2023}:
\begin{eqnarray}
    Q&=&\frac{18(1+2\alpha)}{(2+\alpha)^2}\,H^2,\\[5pt]
    k\rho &=&\frac{f(Q)}{2}-\frac{18(1+2\alpha)}{(2+\alpha)^2}\,H^2f_Q,\label{em1}\\[5pt]
    kp_x&=&-\,\frac{f(Q)}{2}+\frac{6}{2+\alpha}f_Q \big[3H^2+\dot{H} \big]+\frac{216(1+2\alpha)}{(2+\alpha)^3}\,H^2\dot{H}f_{QQ},\label{em2}\\[5pt]
    kp_y&=&-\,\frac{f(Q)}{2}+3f_Q\Big(\frac{1+\alpha}{2+\alpha}\Big)\Big[3H^2+\dot{H}\Big]+\frac{108(1+\alpha)(1+2\alpha)}{(2+\alpha)^3}\,H^2\dot{H}f_{QQ}\label{em3}.
\end{eqnarray}
Again, from the relation between the directional Hubble parameters in equation 
\eqref{H_directional}, the relation between directional scale factors can be 
obtained as
\begin{equation}
a_{1} = c\, a_2^\alpha,
\end{equation}
where $c$ is a constant and for simplification we have taken $c=1$. Thus, 
the average scale factor takes the form $a = a_2^{(\alpha+2)/3}$ and 
hence the total energy density of the Universe can be found as 
 \begin{equation}
 \rho = \rho_{0} {a_{2}}^{-(1+\omega)(2+\alpha)},
 \end{equation}
which follows from the energy-momentum conservation in the studied setting as 
shown in Ref.~\cite{deepjc}. The model-independent generalised field equations 
\eqref{em1}, \eqref{em2} and \eqref{em3} along with other equations derived in 
this section will be used by us in our next sections.

\section{Dynamical system analysis of LRS-BI Universe}\label{sec4}

In this section, we consider two $f(Q)$ gravity models to do the dynamical 
system analysis for the LRS-BI Universe. Through this analysis, we want to 
check the stability of the system by analysing the critical points and the 
nature of those critical points.

\subsubsection{$f(Q) = -\,(Q + 2\Lambda)$ model}
This model is just the extension of the classical $f(Q) = -\,Q$ model, which 
produces classical GR results in $f(Q)$ gravity theory \cite{Sarmah_2023,Q6}. Here the 
additional cosmological term is introduced to get more realistic results. In 
FLRW metric it gives us the $\Lambda$CDM results \cite{Sarmah_2023}. The Friedmann 
equation \eqref{em1} for the given model in terms of density parameters takes 
the form:
\begin{equation}\label{density_friedmann}
\Omega_m  + \Omega_r  + \Omega_{\Lambda} + \Omega_{\sigma} = 1.
\end{equation}
Here, $\Omega_m = k\rho_{m}/3H^2$, $\Omega_r = k\rho_r/3H^2$, 
$\Omega_{\Lambda} = \Lambda/3H^2$ and $\Omega_{\sigma} = \sigma^2/3H^2$ are 
the density parameters of matter, radiation, vacuum energy and anisotropy 
respectively. Using equations \eqref{sigma} and \eqref{H_directional}, the 
density parameter of anisotropy can further be written as 
 \begin{equation}\label{omegasigma}
 \Omega_{\sigma} = \left(\frac{\alpha -1}{\alpha+2}\right)^{2}\!.
\end{equation}  
With this form of $\Omega_{\sigma}$ the equation \eqref{density_friedmann} can 
be written as
\begin{equation}\label{friedmann_alpha}
\Omega_m  + \Omega_r  + \Omega_{\Lambda} + \left(\frac{\alpha -1}{\alpha+2}\right)^{2} = 1.
\end{equation}
Further, for the notational convenience, we denote the density parameters as 
$x = \Omega_m$, $y = \Omega_r$ and $z =\Omega_{\Lambda}$. Hence, with 
these notations the Friedmann equation \eqref{friedmann_alpha} takes the form:
\begin{equation}\label{friedman_new}
x+y+z+\left(\frac{\alpha -1}{\alpha+2}\right)^{2} = 1.
\end{equation}
To obtain the dynamical system of equations for the model, we calculate the 
derivative of $x$ and $y$ with respect to $N = \log {a}$, which is a 
dimensionless time variable. These derivatives of $x$ and $y$ are obtained as
\begin{eqnarray} \label{ds1}
x'=\frac{dx}{dN} =x \Big[(\alpha +2) x+\frac{4}{3} (\alpha +2) y+\frac{(\alpha -1)^2}{(\alpha +2)^2}-(\alpha +2)+\frac{(\alpha -1)^2-3 (\alpha +2)+(\alpha +2) (4-\alpha )}{\alpha +2}\bigg],\\[5pt]
\label{ds2}
y' = \frac{dy}{dN} = y \Big[(\alpha +2) x+\frac{4}{3} (\alpha +2) y+\frac{(\alpha -1)^2}{(\alpha +2)^2}-(\alpha +2)+\frac{(\alpha -1)^2-4 (\alpha +2)+(\alpha +2) (4-\alpha )}{\alpha +2}\bigg],
\end{eqnarray}
where the prime denotes the derivative with respect to $N$.
Again, using equation (\ref{eom2}) for the considered $f(Q)$ model, we obtain 
\begin{equation}\label{hdot}
\frac{\dot{H}}{H^2} = -\left(\frac{\alpha+2}{6}\right) y + \frac{(\alpha+2)}{2}z -\frac{(\alpha -1)^2+(4-\alpha)(\alpha+2)}{2(\alpha+2)}.
\end{equation}
and using equation \eqref{friedman_new} we can express equation \eqref{hdot} 
in terms of variables $x$ and $y$ as
\begin{equation}
\frac{\dot{H}}{H^2} = -\left(\frac{\alpha+2}{2}\right) x - \frac{2(\alpha+2)}{3}y -\frac{(\alpha +2)^2-3(\alpha-1)^2-(4-\alpha)(\alpha+2)}{2(\alpha+2)}.
\end{equation}
The effective equation of state ($\omega_eff$) and deceleration parameter 
($q$) can be written  as
\begin{equation}
\omega_{eff} = -\left(1+ \frac{2 \dot{H}}{3H^2}\right)= \left(\frac{\alpha+2}{3}\right)x + \frac{4(\alpha + 2)}{9}y + \frac{(\alpha + 2)(\alpha -1) - (\alpha + 2)^2 - 3(\alpha -1)^2}{3(\alpha + 2)},\\
\end{equation}
\begin{equation}
 q = -\left(1 + \frac{\dot{H}}{H^2}\right) = \left(\frac{\alpha+2}{2}\right) x + \frac{2(\alpha+2)}{3}y+ \frac{(\alpha + 2)^2 -3(\alpha -1)^2-(\alpha + 2)(6 - \alpha)}{2(\alpha+2)}.
\end{equation}

\begin{figure}[!h]
\centerline{
  \includegraphics[scale = 0.4]{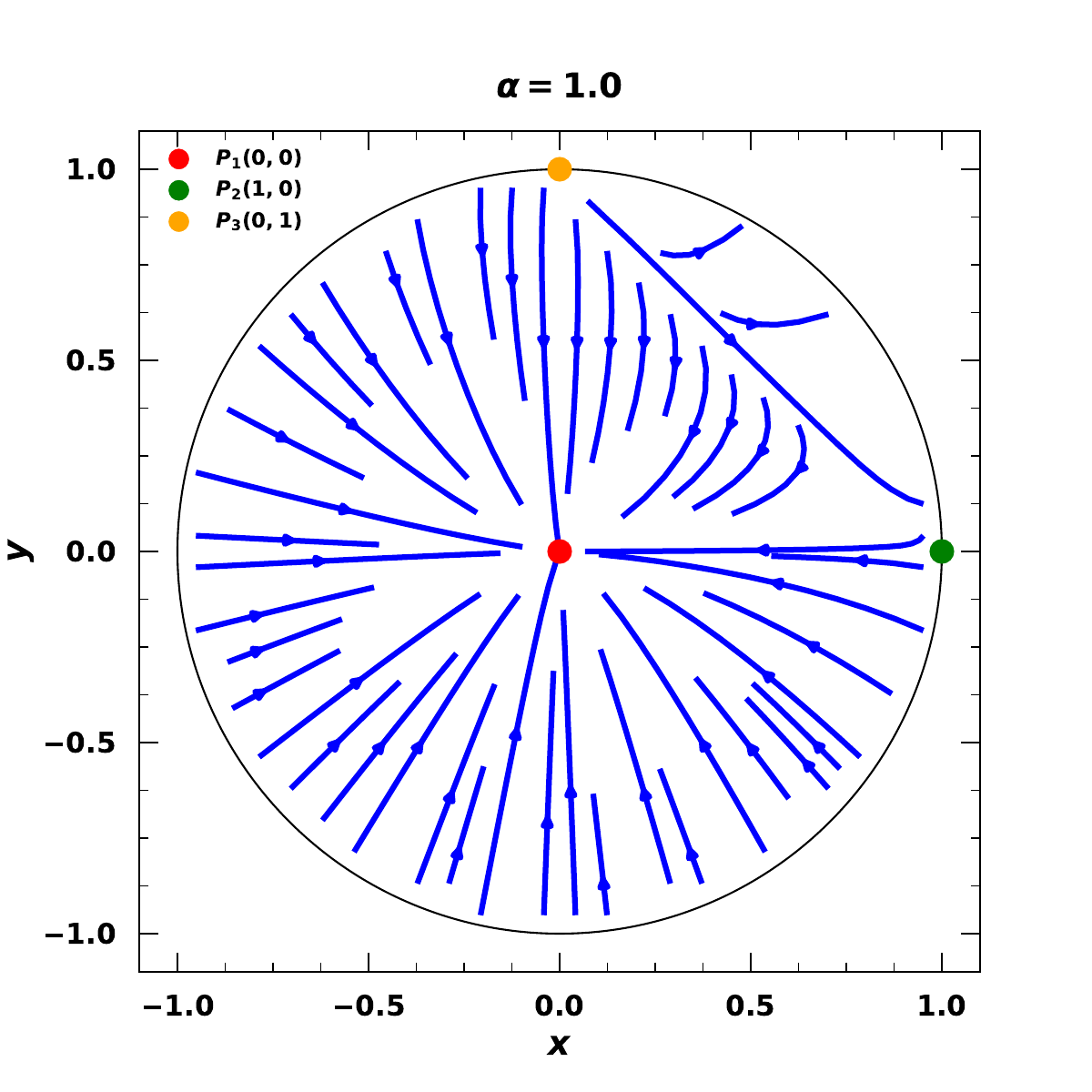}\hspace{0.25cm}
  \includegraphics[scale = 0.4]{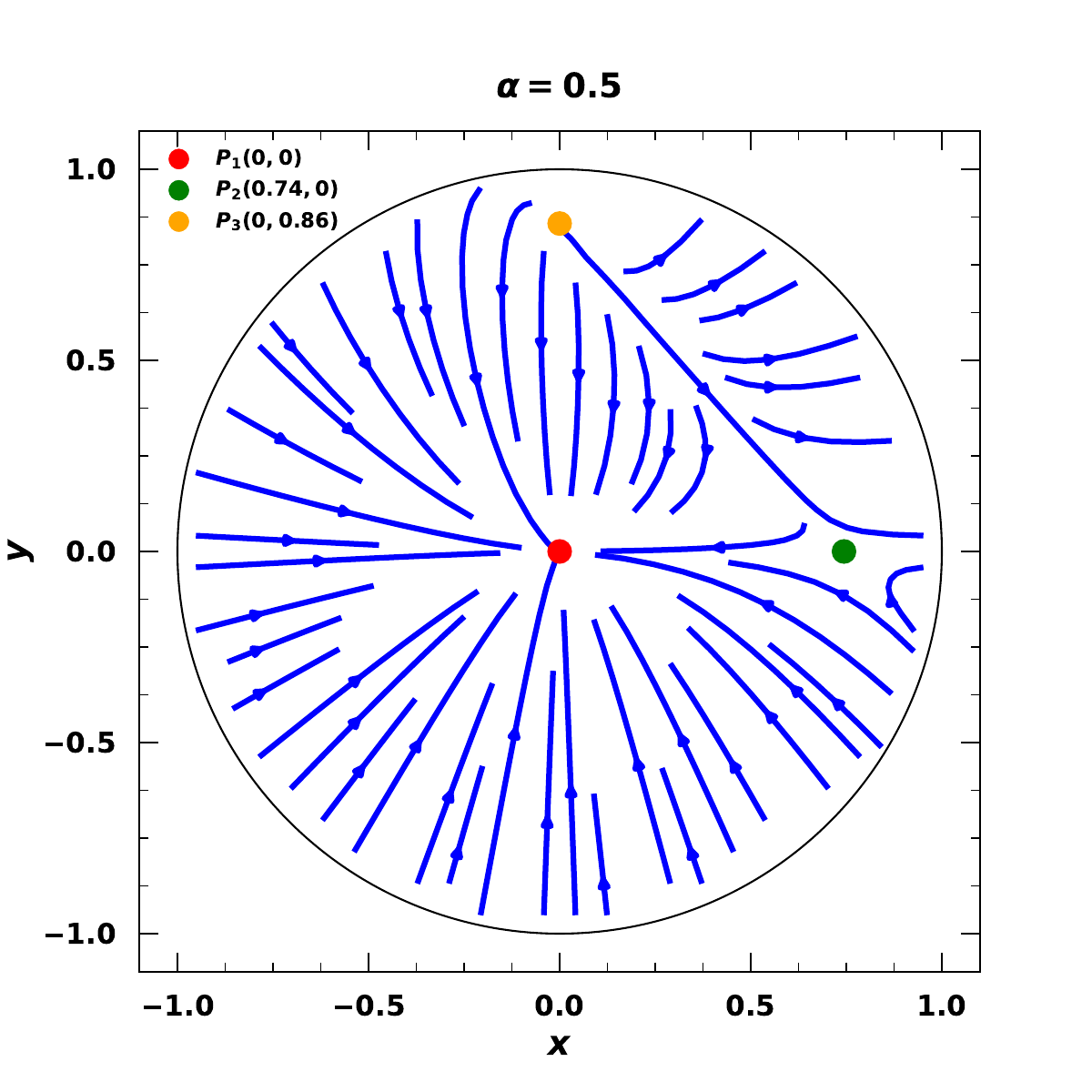}}
 \centerline{
  \includegraphics[scale = 0.4]{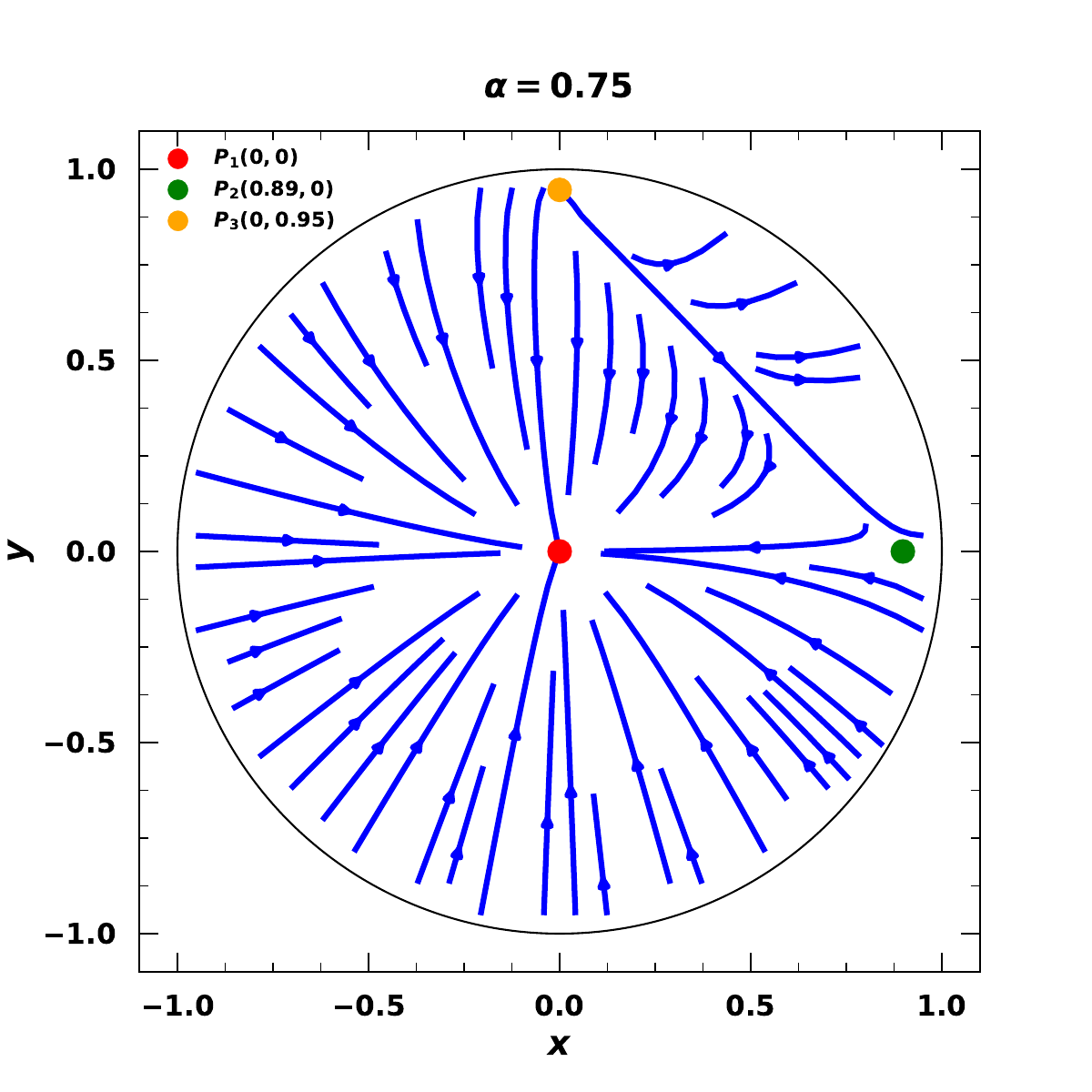}\hspace{0.25cm}
  \includegraphics[scale = 0.4]{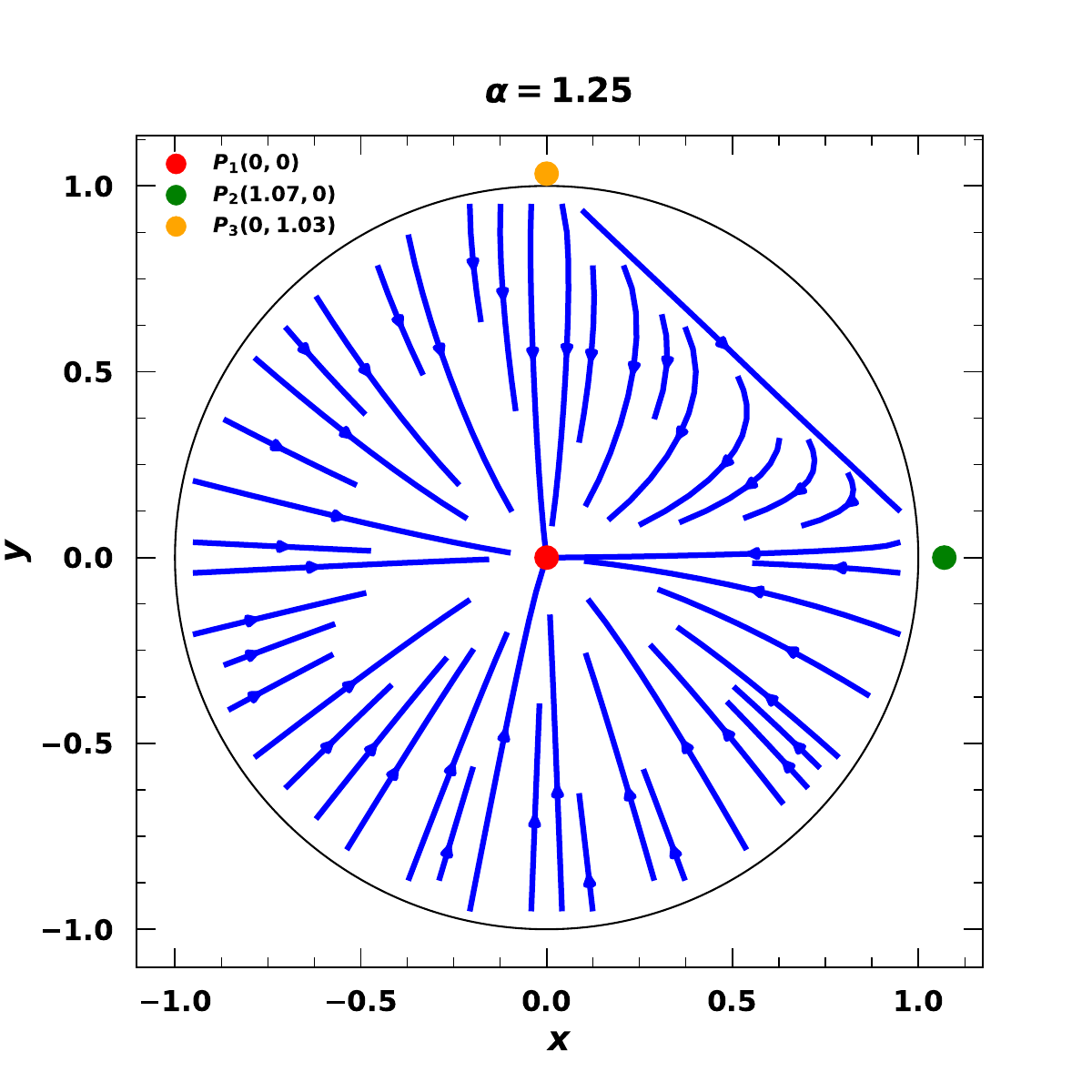}}
  \vspace{-0.2cm}
  \caption{Phase space portraits for $f(Q) = -\,(Q+2\Lambda)$ model for various 
values of $\alpha$ within the constrained range ($0.5 - 1.25$).}
\label{fig1}
\vspace{-1cm}
\end{figure}

The results of the fixed point analysis of the above dynamical system 
(equations \eqref{ds1}, \eqref{ds2}) are shown in Table \ref{table1}. 
Obviously, these fixed point solutions are anisotropic parameter $\alpha$ 
depend. The detailed process of constraining $\alpha$ is discussed in 
Ref.~\cite{Sarmah_2023}, 
which constrained $\alpha$ in between $0.5 - 1.25$ to be consistent with the 
observational data. The phase portraits for the considered model with the 
various $\alpha$ values that include these fixed points are shown in 
Fig.~\ref{fig1}. For $\alpha = 1.0$, the obtained phase portrait is nothing 
but that for the standard $\Lambda$CDM model. However, for other values of 
$\alpha < 1$, the effect of anisotropy shifts the critical points from the 
standard cosmology, which indicates the level of the contribution of anisotropy 
in the total density of the Universe. Further, for $\alpha > 1$, the system 
provides the solutions that are physically impossible as the values of the 
density parameters $x$ and $y$ can not be greater than $1$. Thus $P_2$ and 
$P_3$ points in Table \ref{table1} give physically consistent results for
$\alpha < 1$ along with some anisotropic effects. All phase space
portraits with $\alpha \leq 1$ show that $P_1$ is a stable point while $P_3$
is a repeller or unstable point and $P_2$ is a saddle point. There is a
heteroclinic path from $P_3 \rightarrow P_2\rightarrow P_1$ indicating the
path the Universe follows while evolving. It indicates that the Universe
undergoes a radiation-dominated phase followed by a matter-dominated phase
and finally reaches to dark energy-dominated phase. Thus in all cases, the
model provides at least a dark energy-dominated phase irrespective of the
value of $\alpha$.
\begin{center}
\begin{table}[!h]
\caption{The fixed point solutions for $f(Q)= -\,(Q+2\Lambda)$ model.}
\vspace{5pt}
\scalebox{0.95}{
\begin{tabular}{|c|c|c|c|c|c|}
\hline 
\rule[1ex]{0pt}{2.5ex} Fixed point &  $(x=\Omega_m,y=\Omega_r)$ &  $z=\Omega_\Lambda$ &  Eigenvalues  & $\omega_{eff}$   &  $q$  \\ 
\hline
\rule[1ex]{0pt}{2.5ex} $P_1 $ & $\Big(0, 0\Big)$ & $\frac{6 \alpha +3}{(\alpha +2)^2}$ & $\Big[\frac{-\alpha ^3-9 \alpha ^2-21 \alpha -5}{(\alpha +2)^2},\frac{-\alpha ^3-8 \alpha ^2-17 \alpha -1}{(\alpha +2)^2}\Big]$ & $\frac{-\alpha ^2+\alpha -3}{\alpha +2}$ & $-\frac{3 \alpha ^2-4 \alpha +7}{2 \alpha +4}$\\ 
\rule[1ex]{0pt}{2.5ex} $P_2$ &$\Big(\frac{\alpha ^3+8 \alpha ^2+17 \alpha +1}{(\alpha +2)^3},0\Big)$ & $-\frac{\alpha ^3+2 \alpha ^2+2 \alpha -5}{(\alpha +2)^3}$ & $\Big[-1,\frac{\alpha ^3+8 \alpha ^2+17 \alpha +1}{(\alpha +2)^2}\Big]$&$\frac{-2 \alpha ^3+5 \alpha ^2+14 \alpha -17}{3 (\alpha +2)^2}$ &$\frac{-2 \alpha ^3+6 \alpha ^2+18 \alpha -13}{2 (\alpha +2)^2}$\\
\rule[1.25ex]{0pt}{2.5ex} $P_3$ &$\left(0,\frac{3 (\alpha ^3+9 \alpha ^2+21 \alpha +5)}{4 (\alpha +2)^3}\right)$ & $-\frac{3 (\alpha ^3+\alpha ^2+\alpha -3)}{4 (\alpha +2)^3}$ & $\Big[1,\frac{\alpha ^3+9 \alpha ^2+21 \alpha +5}{(\alpha +2)^2}\Big]$ & $\frac{-2 \alpha ^3+6 \alpha ^2+18 \alpha -13}{3 (\alpha +2)^2}$ &$\frac{-2 \alpha ^3+7 \alpha ^2+22 \alpha -9}{2 (\alpha +2)^2}$\\ 
\hline
\end{tabular}
}
\label{table1}
\end{table} 
\end{center}

\begin{figure}[!h]
\centerline{
  \includegraphics[scale = 0.4]{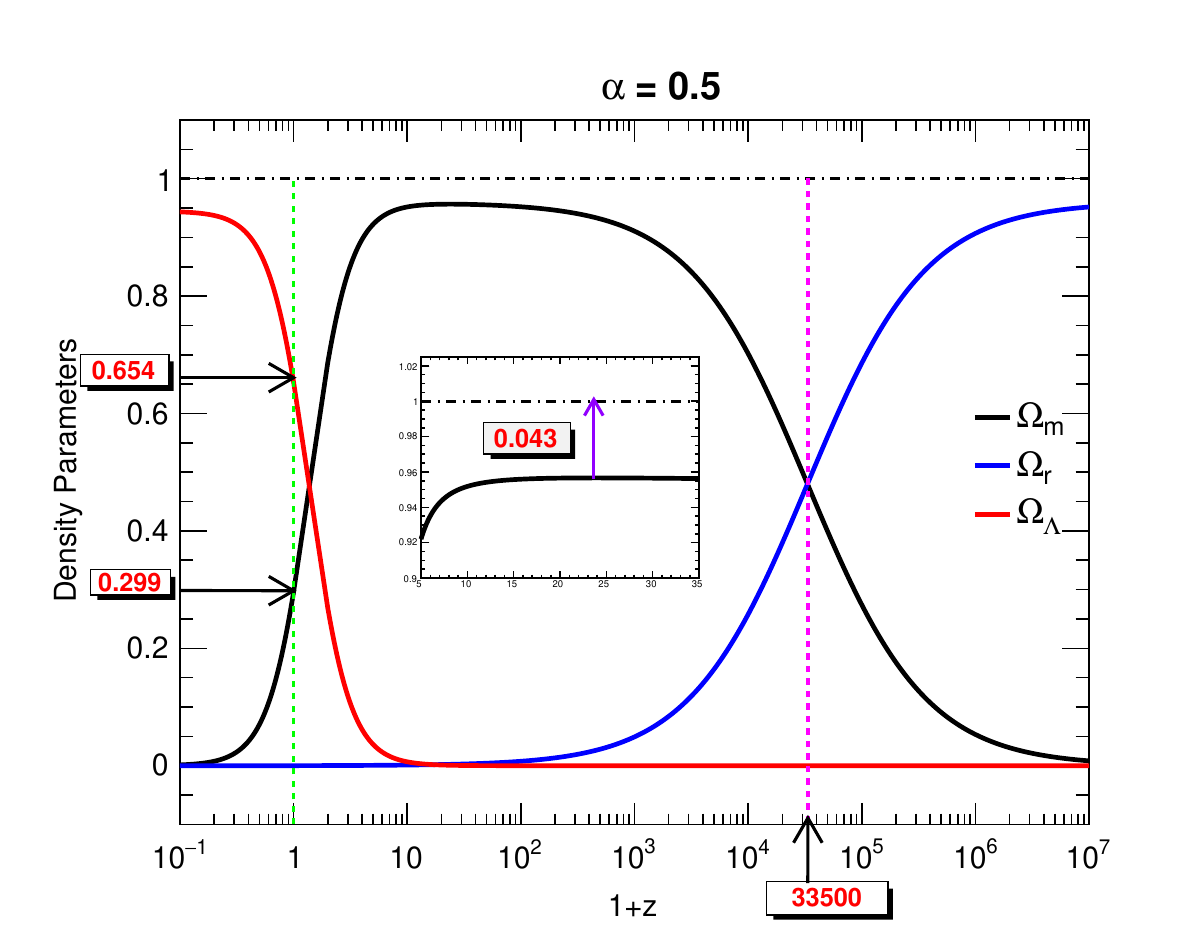}\hspace{0.0cm}
  \includegraphics[scale = 0.4]{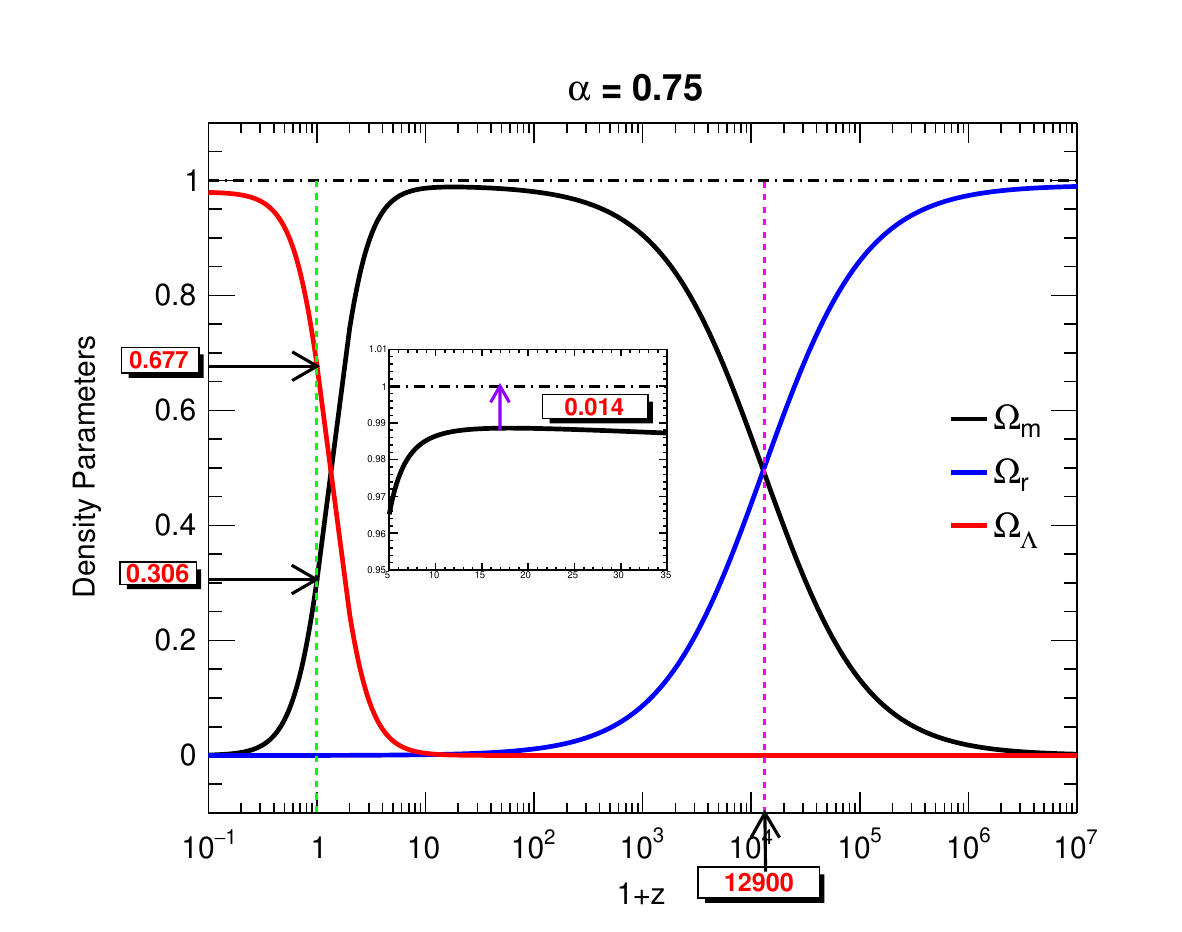}}
  \centerline{
  \includegraphics[scale = 0.4]{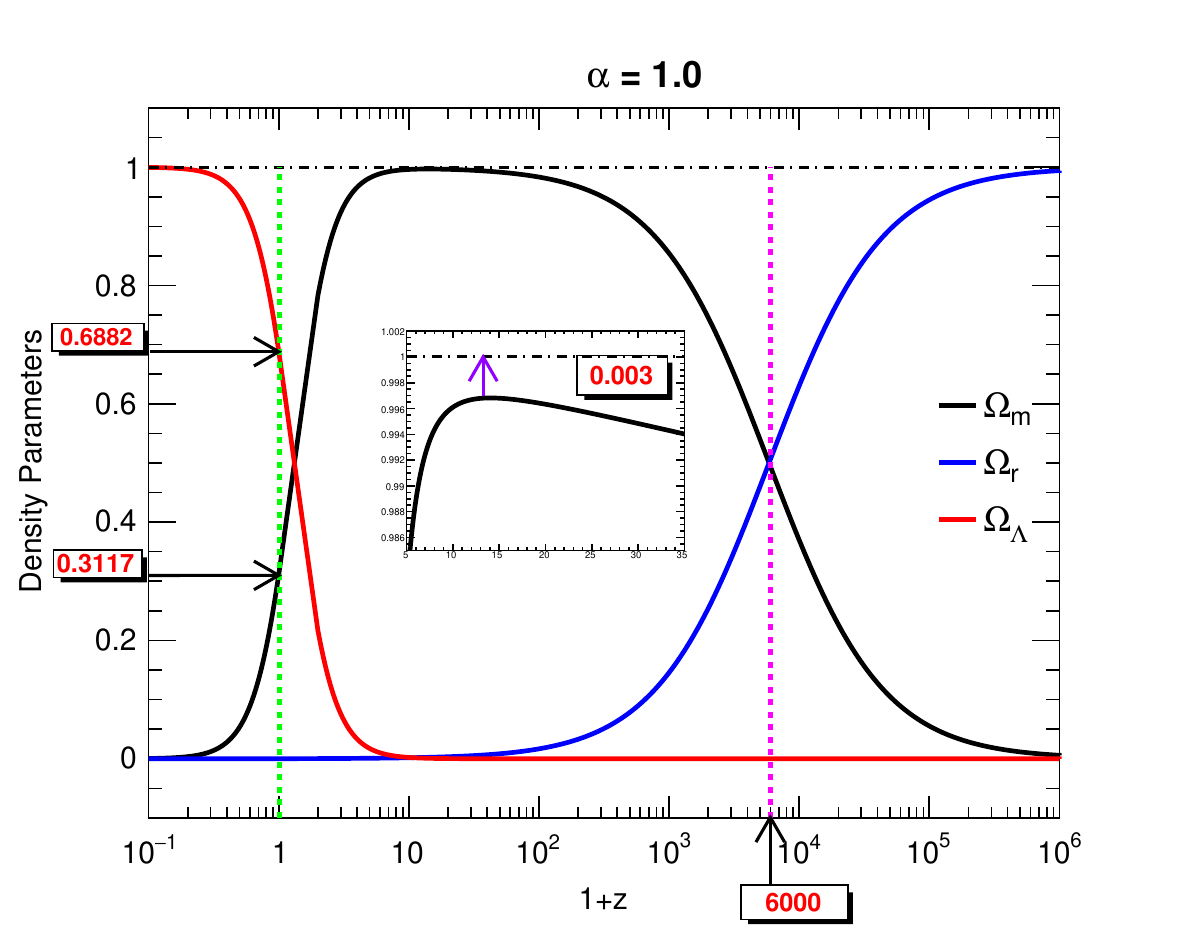}
  }
\caption{Behaviours of density parameters with respect to redshift $z$ for 
$\alpha = 0.5$, $0.75$ and $1.0$.} 
\label{fig_2}
\end{figure}

Moreover, we plot the density parameters of matter ($\Omega_{m}$), radiation 
($\Omega_{r}$) and dark energy ($\Omega_{\Lambda}$) against cosmological 
redshift ($z$) in Fig.~\ref{fig_2} for $\alpha = 0.5$, $0.75$ and $1.0$. Here 
we restrict our analysis for $\alpha \leq 1$ as the phase portrait analysis 
shows that for $\alpha > 1$ the system is physically inconsistent.
From Fig.~\ref{fig_2}, we see that the density parameters never reached $1$ 
for $\alpha \leq 1$, which is shown in the inset for $\alpha = 0.5$, 
$0.75$ and $1.0$ plots. The $\alpha = 1$ plot gives the standard $\Lambda$CDM 
results. For $\alpha = 0.5$ the maximum value of $\Omega_{m}$ is $0.956818$ 
within the range of $z =$ ($22.7227 - 23.2232$). For $\alpha = 0.75$ the 
maximum value is $0.988569$, which lies within the $z$ range 
($16.7167 - 17.2172$). For the isotropic standard $\Lambda$CDM model 
(i.e.~for $\alpha =1$), the maximum value of the density parameter of 
matter is $0.996818$ within $z = (12.9129 - 13.2132)$. The gap between the 
unity and the maximum value reached by the density parameter of matter 
indicates that there are some contributions from other density parameters like 
radiation and dark energy for the standard $\Lambda$CDM model. But for 
$\alpha = 0.5$ and $\alpha =0.75$ plots, the gap is much wider and it is not 
only due to some contributions coming from matter and radiation but there is 
some contribution of anisotropy also present here. By using equation 
\eqref{omegasigma} we can calculate the value of the anisotropy parameter for 
$\alpha = 0.5$ and $\alpha =0.75$ as $0.04$ and $0.00826$ respectively, and 
this is also reflected in the plots of Fig.~\ref{fig1}. This anisotropic 
contribution of energy stored in the form of shear arises in spacetime due to 
different expansion rates at different directions of the Universe.

Also, due to the anisotropy, the maximum value i.e.~peak of density parameter 
of matter is shifted from the original $\Lambda$CDM result as the value of 
$\alpha$ decreases from the unity which we discussed a little earlier. Thus, 
we can say that the greater the anisotropy, the greater the shift of peak from 
the standard result. Therefore the maximum matter density contribution period 
of the Universe for higher anisotropy is earlier than the standard cosmology. 
Further, the anisotropic effect also changes the length of the matter 
domination era as compared to the isotropic case. The redshift value $z$ for 
which the matter density had started dominating the radiation density for 
$\alpha = 0.50$ is around $33499$ and for $\alpha = 0.75$ this value of $z$ is 
around $12899$. However, for the isotropic case, the value of $z$ for which 
the matter started dominating over radiation is around $5999$. From these 
observations, we can say that the anisotropy results in the elongated 
matter-dominated era as compared to an isotropic  Universe. It is to be noted 
that we use Planck 2018 data \cite{Planck_2018} for our calculations.
   
Based on the above discussion, we can say that the simple 
$f(Q) = -\,(Q+2 \Lambda)$ model provides some interesting results in 
cosmological studies when considering anisotropy.

\subsubsection{$f(Q) = -\beta Q^{n}$ model}
This is a power-law form of model with constant multiplier $\beta$ and $n$ be 
the exponent. This is one of the simplest but powerful models, which is 
consistent with the current observational cosmological data from different 
data sources up to a certain extent. For this model, the Friedmann equation 
\eqref{em1} takes the form:
\begin{equation}\label{friedmann_power}
\frac{\big(x+y+z\big)^{\frac{1}{n}}}{2\big[\beta(n-\frac{1}{2})^{\frac{1}{n}} 3^{n-1} H^{2n-2}\big]}+ \Omega_{\sigma} = 1.
\end{equation}
Here $x$, $y$ and $z$ are the dynamical variables as mentioned in the previous 
model case. We calculate the derivatives of these variables with respect to 
$N$, which are obtained as
\begin{align}\label{dy_power1}
x' = x\bigg[ \left(\frac{(2+\alpha)}{n}\right)\left(\frac{4}{3}\,y+x\right)-\frac{2(2+\alpha)}{n}\left\{\beta \big(n-\frac{1}{2}\big)^{\frac{1}{n}}3^{n-1}H^{2n-2}-\frac{(n-1)}{(2n-1)}\right\}\left(\frac{\alpha -1}{\alpha+2}\right)^{2}\nonumber \\[5pt] +\,\frac{6n-2-\alpha-3n(2n-1)}{n(2n-1)}-\frac{2(2+\alpha)}{n}\left\{\beta \big(n-\frac{1}{2}\big)^{\frac{1}{n}}3^{n-1}H^{2n-2}\right\}\bigg],
\end{align}
\begin{align}\label{dy_power2}
y' = y\bigg[\left(\frac{(2+\alpha)}{n}\right)\left(\frac{4}{3}\,y+x\right)-\frac{2(2+\alpha)}{n}\left\{\beta \big(n-\frac{1}{2}\big)^{\frac{1}{n}}3^{n-1}H^{2n-2}-\frac{(n-1)}{(2n-1)}\right\}\left(\frac{\alpha -1}{\alpha+2}\right)^{2}\nonumber \\[5pt] +\,\frac{6n-2-\alpha-4n(2n-1)}{n(2n-1)}-\frac{2(2+\alpha)}{n}\left\{\beta \big(n-\frac{1}{2}\big)^{\frac{1}{n}}3^{n-1}H^{2n-2}\right\}\bigg].
\end{align}
Further, we can calculate
\begin{align}\label{Hdot}
\frac{\dot{H}}{H^2} = -\frac{2+\alpha}{2n}\left[\frac{4}{3}y+x-\left\{2^n \beta^n (n-\frac{1}{2})3^{n(n-1)} H^{n(2n-2)}\right\}\left(\frac{2\alpha+1}{\alpha+2}\right)^n\left(\frac{3}{\alpha+2}\right)^n\right]\nonumber \\[5pt] -\,\frac{2+\alpha}{n}\big(\frac{n-1}{2n-1}\big)\big(\frac{\alpha -1}{\alpha+2}\big)^{2}+\frac{6n-\alpha-2}{2n(2n-1)}.
\end{align}
The effective equation of state ($\omega_{eff}$) and deceleration parameter 
($q$) for this model can be written respectively by using the equation 
\eqref{Hdot} as
\begin{align}\label{omfp}
\omega_{eff} = \frac{2(2+\alpha)}{6n}\left[\frac{4}{3}y+x-\left\{2^n \beta^n (n-\frac{1}{2})3^{n(n-1)} H^{n(2n-2)}\right\}\big(\frac{2\alpha+1}{\alpha+2}\big)^n\big(\frac{3}{\alpha+2}\big)^n\right]\nonumber\\[5pt] +\, \frac{2+\alpha}{n}\big(\frac{n-1}{2n-1}\big)\big(\frac{\alpha -1}{\alpha+2}\big)^{2} + \frac{2(6n-\alpha-2)-6n(2n-1)}{6n(2n-1)},
\end{align}
\begin{align}\label{qp}
q = \frac{2+\alpha}{2n}\left[\frac{4}{3}y+x-\left\{2^n \beta^n (n-\frac{1}{2})3^{n(n-1)} H^{n(2n-2)}\right\}\big(\frac{2\alpha+1}{\alpha+2}\big)^n\big(\frac{3}{\alpha+2}\big)^n\right]\nonumber\\[5pt] +\,\frac{2+\alpha}{n}\big(\frac{n-1}{2n-1}\big)\big(\frac{\alpha -1}{\alpha+2}\big)^{2}+\frac{6n-\alpha-2-2n(2n-1)}{2n(2n-1)}.
\end{align}

Now, as in the previous case if we assume that the total mass-energy density 
of the Universe consists of $\Omega_m, \Omega_r, \Omega_\Lambda$ and 
$\Omega_\sigma$, which satisfy equation \eqref{density_friedmann},
then from equation \eqref{friedmann_power} we can write,
\begin{equation}
H^{2n-2} = \frac{\left\{3(1-\Omega_\sigma)\right\}^{1-n}}{2^n (n-\frac{1}{2})\beta}.
\end{equation}
Again with the definition of $\Omega_{\sigma}$ from equation 
\eqref{omegasigma}, equations \eqref{dy_power1}, \eqref{dy_power2}, 
\eqref{omfp} and \eqref{qp} can be written in the forms:
\begin{align}
x'= \frac{x}{n} \bigg[\frac{\alpha +6 n^2-9 n+2}{1-2 n}-6^{1-n}(\alpha +2)\Big(n-\frac{1}{2}\Big)^{\frac{1}{n}-1} \Big(\frac{2 \alpha +1}{(\alpha +2)^2}\Big)^{1-n} \beta ^{\frac{1}{n}-1} - 2(\alpha+2)^{-1}(\alpha -1)^2\nonumber\\[5pt] \times \left\{(2^{-n} \Big(n-\frac{1}{2}\Big)^{\frac{1}{n}-1} \Big(\frac{6 \alpha +3}{(\alpha +2)^2}\Big)^{1-n} \beta ^{\frac{1}{n}-1}+\frac{1-n}{2 n-1}\right\} +(\alpha +2) \Big(x+\frac{4 y}{3}\Big)\bigg],
\end{align}

\begin{align}
y'=\frac{y}{n} \bigg[\frac{\alpha +8 n^2-10 n+2}{1-2 n} -6^{1-n}(\alpha +2)\Big(n-\frac{1}{2}\Big)^{\frac{1}{n}-1} \Big(\frac{2 \alpha +1}{(\alpha +2)^2}\Big)^{1-n} \beta ^{\frac{1}{n}-1} - 2(\alpha+2)^{-1}(\alpha -1)^2\nonumber\\[5pt]\times \left\{2^{-n} \Big(n-\frac{1}{2}\Big)^{\frac{1}{n}-1} \Big(\frac{6 \alpha +3}{(\alpha +2)^2}\Big)^{1-n} \beta ^{\frac{1}{n}-1}+\frac{1-n}{2 n-1}\right\} + (\alpha +2) \Big(x+\frac{4 y}{3}\Big)\bigg],
\end{align}

\begin{align}
\omega_{eff} = -1-\frac{2}{3}  \left[\frac{(\alpha -1)^2 (n-1) n}{2 (\alpha +2) (2 n-1)}-\frac{-\alpha +6 n-2}{2 n (2 n-1)}\right]+\frac{1}{3} (\alpha +2) n \bigg\{-2 \left(1-\frac{(\alpha -1)^2}{(\alpha +2)^2}\right) 3^{(n-1) n}\beta \nonumber\\[5pt] \times\left(n-\frac{1}{2}\right)\bigg\}\left\{ \left(2^{-n} 3^{1-n} \beta^{-1} \left(n-\frac{1}{2}\right)^{-1} \left(1-\frac{(\alpha -1)^2}{(\alpha +2)^2}\right)^{1-n}\right)^n\!\!+x+\frac{4 y}{3}\right\},
\end{align}
\begin{align}
q=-1+\frac{(\alpha -1)^2 (n-1) n}{2 (\alpha +2) (2 n-1)}-\frac{-\alpha +6 n-2}{2 n (2 n-1)}+\frac{1}{2} (\alpha +2) n \bigg\{-2 \left(1-\frac{(\alpha -1)^2}{(\alpha +2)^2}\right) 3^{(n-1) n}\beta\nonumber\\[5pt]\times \left(n-\frac{1}{2}\right)\bigg\}\left\{\left(2^{-n} 3^{1-n} \beta^{-1}  \left(n-\frac{1}{2}\right)^{-1} \left(1-\frac{(\alpha -1)^2}{(\alpha +2)^2}\right)^{1-n}\right)^n\!\!+x+\frac{4 y}{3}\right\}.
\end{align}
\begin{center}
\begin{table}[b]
\caption{The fixed points numerical solutions for $f(Q)= -\beta Q^n$ model for
$n = 1$.}
\vspace{5pt}
\begin{tabular}{|c|c|c|c|c|c|c|c|}
\hline
\rule[-1ex]{0pt}{2.5ex}$\alpha$& Fixed point&  $(x = \Omega_m,y= \Omega_r)$ &$z=\Omega_{\Lambda}$&  Eigenvalues  &  $\omega_{eff}$   &  $q$  & Remarks \\
\hline

\rule[2ex]{0pt}{2.5ex} &$ P_{a11}$  &$\left(0,0 \right)$ & $1$ & $\left[-\,4,-\,3\right]$& $-\,1$ & $-\,1$& \\

\rule[2ex]{0pt}{2.5ex} $\alpha = 1 $ &$ P_{a12}$  &$\left(1, 0 \right)$ & $0$ & $\left[-\,1,3\right]$& $0$ & $\frac{1}{2}$ &$\Lambda$CDM \\

\rule[2ex]{0pt}{2.5ex} &$ P_{a13}$  &$\left(0,1 \right)$ & $0$ & $\left[1,4\right]$& $\frac{1}{3}$ & $1$ &\\

\rule[-1ex]{0pt}{2.5ex} & &&& &  & &\\
\hline

\rule[2ex]{0pt}{2.5ex} &$P_{a21}$&$\left( 0,0\right)$  & $0.96$ & $\left[-\,3.1,-2.1\right]$&$-\,0.63$ & $-\,0.45$ &\\

\rule[2ex]{0pt}{2.5ex}$\alpha = 0.5$&$P_{a22}$&$\left(0.84,0\right)$&$0.12$&$\left[2.1,-\,1.0\right]$ & $0.07$&$0.6$& Anisotropic\\

\rule[2ex]{0pt}{2.5ex}&$P_{a23}$&$\left(0,0.93\right)$ &$0.03$& $\left[3.1,1.0\right]$ & $0.4$ & $1.1$& \\

\rule[-1ex]{0pt}{2.5ex} & &&& &  & &\\
\hline

\rule[2ex]{0pt}{2.5ex} &$P_{a31}$&$\left( 0,0\right)$  & $0.99$ & $\left[-\,3.52,-\,2.52\right]$&$-\,0.83$&$-\,0.74$&\\

\rule[2ex]{0pt}{2.5ex} $\alpha = 0.75$&$P_{a32}$&$\left(0.92, 0\right),$&$0.07$&$\left[2.52,-\,1.0\right]$ & $0.015$ & $0.52$ & Anisotropic\\

\rule[2ex]{0pt}{2.5ex}&$P_{a33}$&$\left(0, 0.97\right)$ &$0.03$& $\left[3.52,1.0\right]$ & $0.35$&$1.02$& \\

\rule[-1ex]{0pt}{2.5ex} & &&& &  & &\\
\hline

\rule[2ex]{0pt}{2.5ex} &$P_{a41}$&$\left( 0,0\right)$  & $0.99$ & $\left[-\,4.52,-\,3.52\right]$&$-\,1.16$&$-\,1.24$&$P_{a42}\, \&\, P_{a43}$\\

\rule[2ex]{0pt}{2.5ex} $\alpha = 1.25$&$P_{a42}$&$\left(1.08, 0\right)$&$-\,0.089$&$\left[3.52,-\,1.\right]$ & $0.012$ & $0.52$ & are \\

\rule[2ex]{0pt}{2.5ex} &$P_{a43}$&$\left( 0,1.04\right)$ &$-\,0.05$ & $\left[4.52,1.0\right]$ & $0.35$ & $1.02$ & unphysical\\
\rule[-1ex]{0pt}{2.5ex} & &&& &  & &\\

\hline
\end{tabular}
\label{table2}
\end{table}
\end{center}

It is evident that the equations derived from this model depend on two model 
parameters $\beta$ and $n$ along with anisotropic parameter $\alpha$. 
As in the case of the parameter $\alpha$ as mentioned earlier, here we use the 
constrained values of parameters $n$ and $\beta$ also within the ranges 
($0.95 - 1.05$) and ($0.75 - 1.5$) respectively that are obtained from the 
Ref.~\cite{Sarmah_2023}. These ranges of values of $n$ and $\beta$ are 
consistent with the observational Hubble data ($H$) and distance modulus 
($D_m$) data and hence are used for the fixed point analysis of this model. 
Results of this analysis have been listed in three tables: Table \ref{table2}, 
Table \ref{table3} and Table \ref{table4} for $ n = 1.0$, $0.95$ and $1.05$ 
respectively. From the Table \ref{table2}, 
one can see that for $n = 1$ and $\alpha = 1$, the system provides the 
$\Lambda$CDM results. For the other values of $\alpha < 1$, the system provides 
solutions with the anisotropic effect and $\alpha > 1$ provides physically 
impossible critical points $P_{a42}$ and $P_{a43}$ as explained in the previous 
model case. It is to be noted that for the $n=1$ situation, the results are 
independent of the model parameter $\beta$. Table \ref{table3}, which shows 
the fixed points for $n = 0.95$, reveals that the fixed points and all the 
cosmological parameters depend upon not only the anisotropic parameter 
$\alpha$ but also on the model parameter $\beta$. Similarly, Table 
\ref{table4} for $n = 1.05$ also shows that fixed points and other cosmological
parameters depend on both $\alpha$ and $\beta$.
\begin{figure}[!h]
\centerline{
  \includegraphics[scale = 0.4]{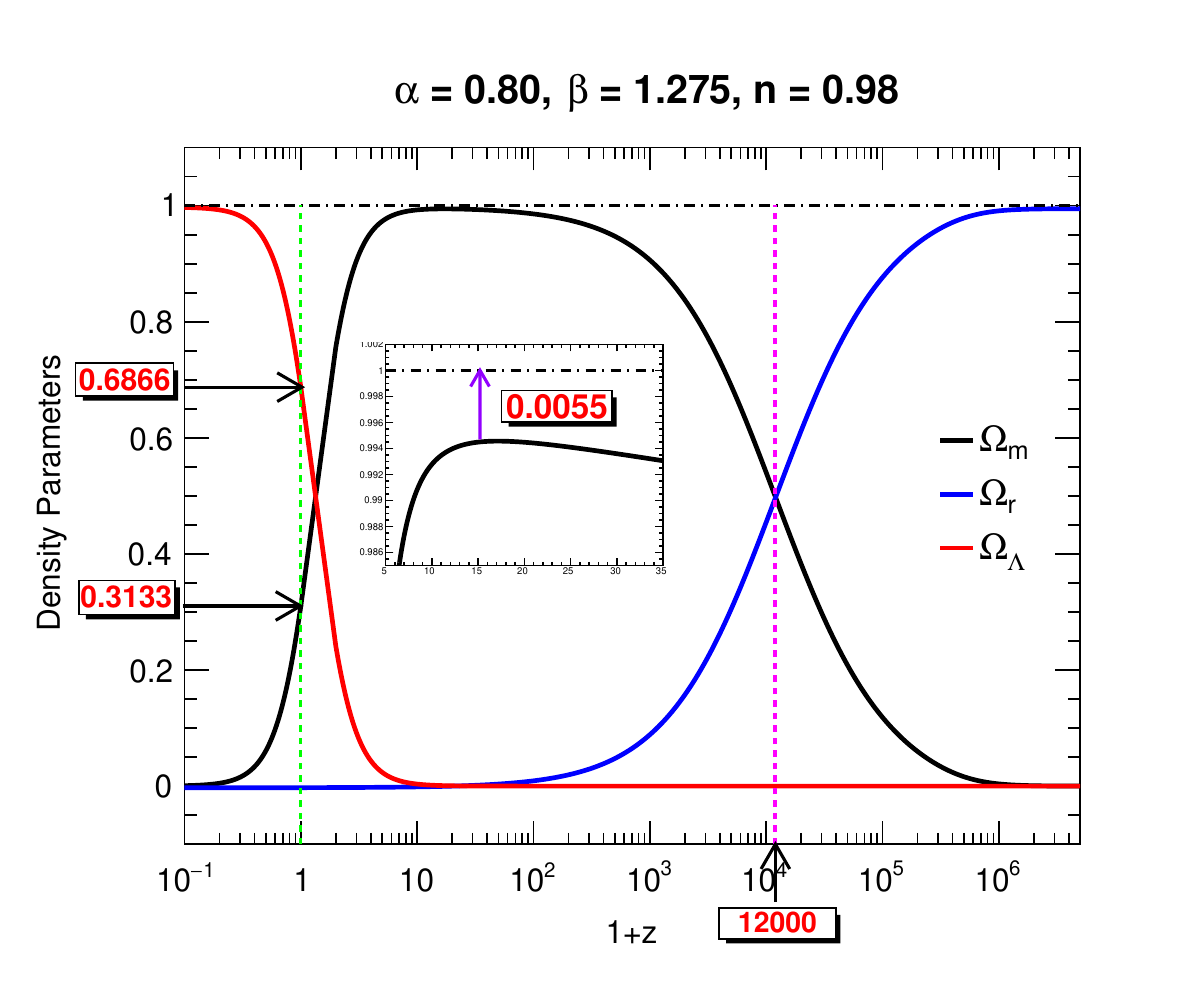}\hspace{0.0cm}
  \includegraphics[scale = 0.4]{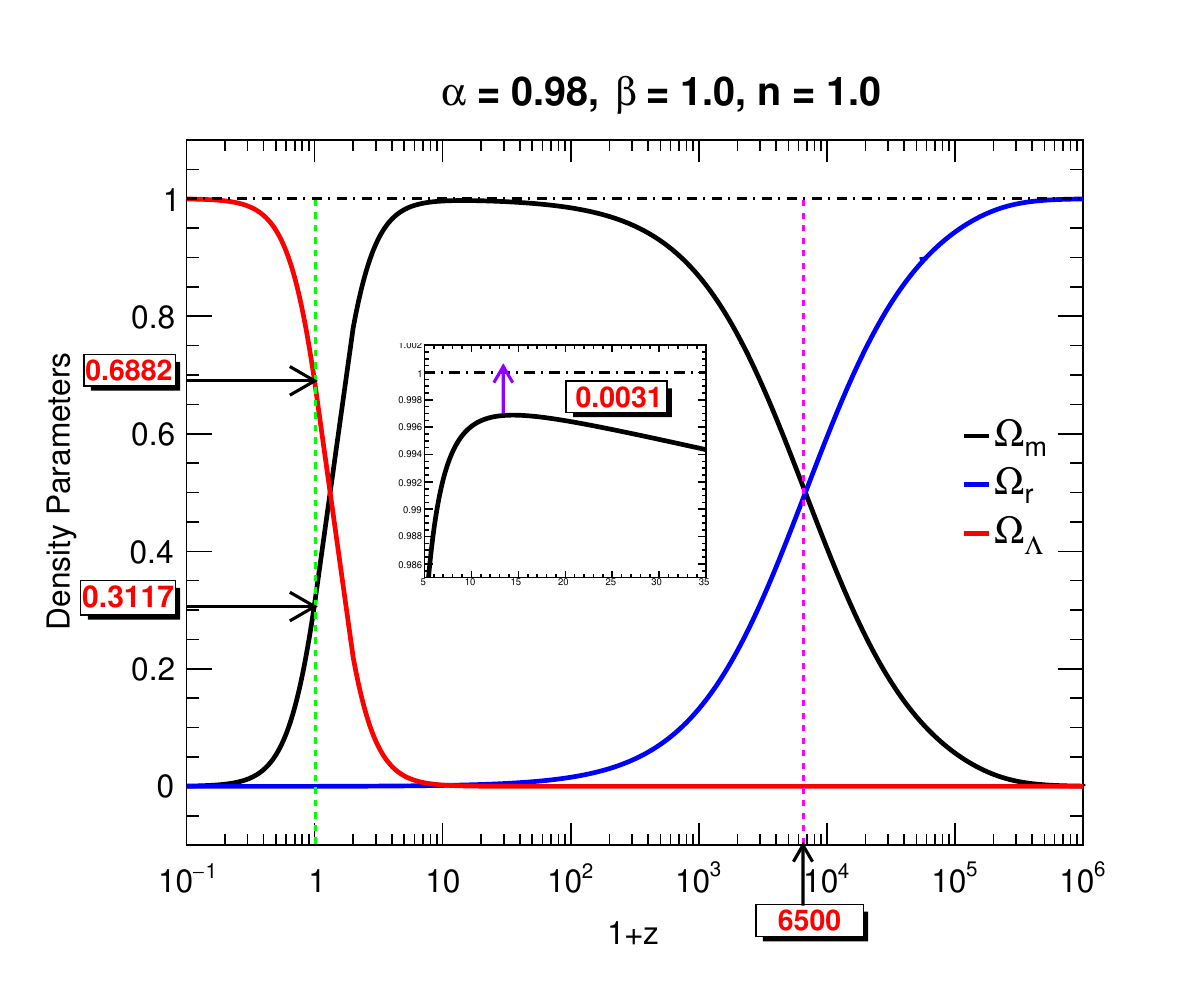}}
\vspace{-0.2cm}
\caption{Density parameters versus cosmological redshift $z$ for two different
sets of model parameters.}
\label{fig3}
\end{figure}

To get a physically viable phase portrait, we further constrain the model 
parameters $\alpha, \beta, n$ by plotting the density parameters against 
redshift for the considered model as shown in Fig.~\ref{fig3}. During our 
study, it is found that for the considered model the range of model parameter 
$\beta$ lies in between ($1 - 1.275$), $n$ lies in between ($0.98 - 1.0$) and 
the anisotropy parameter $\alpha$ lies in between ($0.8 - 0.98$). Here again, 
it is found that the density parameters like $\Omega_{m}$ never reached unity 
due to the anisotropy and some other contributions from radiation and dark 
energy as we discussed in our previous model case. Further, the anisotropy 
affects the length of the matter-dominated era and it is found that the $z$ 
value for which the matter density starts to exceed the radiation density for 
$\alpha = 0.8$ is at $11999$ and for $\alpha = 0.98$, its value is $6499$. 
Hence, the presence of anisotropy may extend the length of the 
matter-dominated era and thus affect the cosmic history of the Universe.

Similar to the previous model, here also the maximum value of the density 
parameter of matter for both the plots shifted from the standard $\Lambda$CDM 
result. For $\alpha = 0.98$, $\beta = 1$ and $n = 1$, the peak is within 
$(12.5125 - 14.6146)$ and for $\alpha = 0.8$, $\beta = 1.275$ and $n = 0.98$ 
the peak is within $(13.7137 - 15.6156)$ as seen in Fig.~\ref{fig3}. Both 
these ranges of $z$ are greater than the standard $\Lambda$CDM result as 
discussed in the previous model. Thus, anisotropy can play some role in the 
evolution of the Universe.

\begin{figure}[!h]
\centerline{
\includegraphics[scale = 0.4]{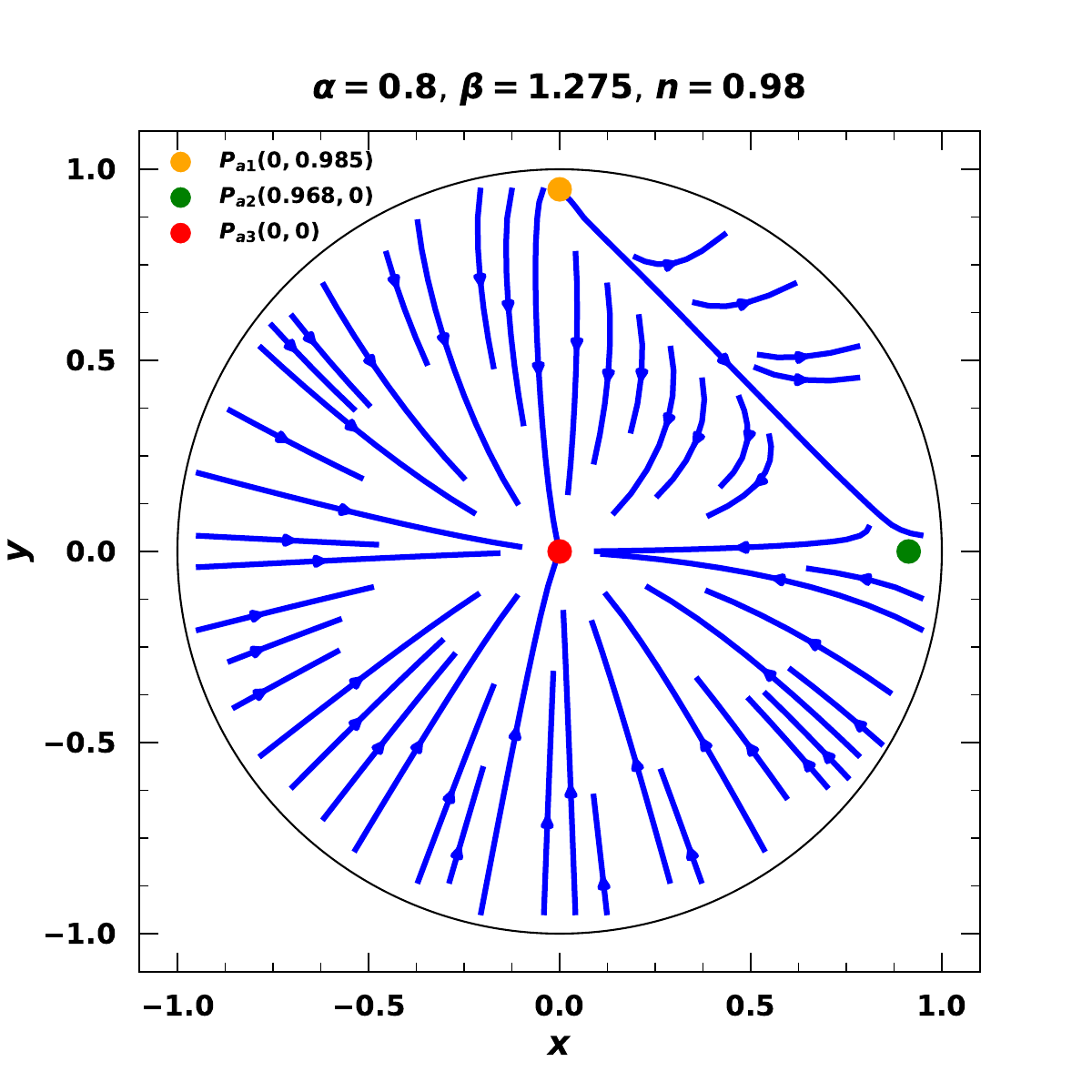}\hspace{0.25cm}
  \includegraphics[scale = 0.4]{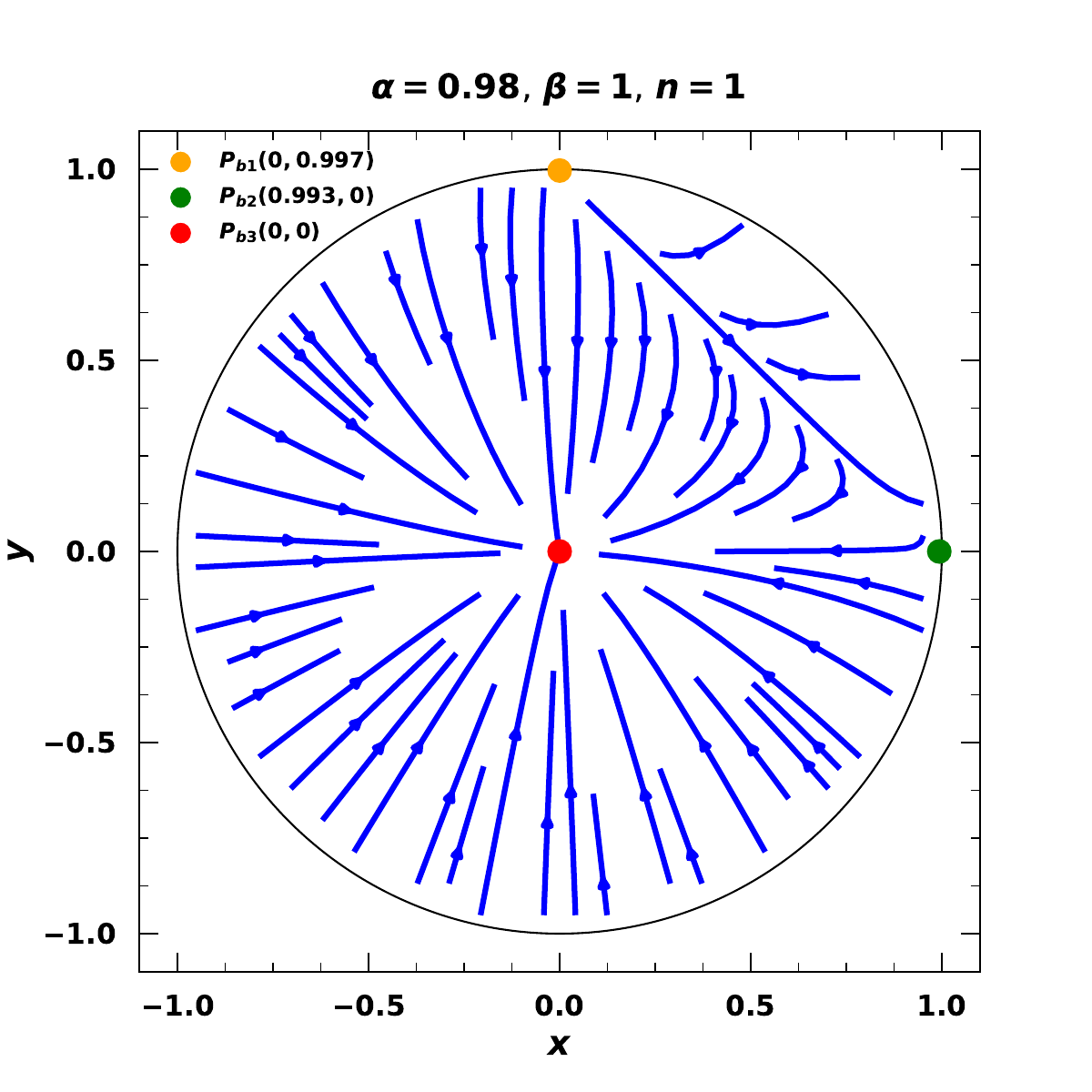}
  }
\vspace{-0.2cm}  
\caption{Phase space portraits for the $f(Q) = \beta Q^{n}$ model with 
($\alpha = 0.8, \beta = 1.275, n = 0.98$) and ($\alpha = 0.98, \beta = 1, 
n = 1$) sets of parameters.} 
\label{fig_4}
\end{figure}

The phase space portrait for the constrained range of model parameters 
$\alpha$, $\beta$ and $n$ from the density parameters versus redshift plots 
have been shown in Fig.~\ref{fig_4}. Like in the previous model, the critical 
points show similar stability and nature, and also provide the heteroclinic 
sequences of radiation-dominated, matter-dominated and dark energy-dominated 
phases through $P_{x1}\rightarrow P_{x2} \rightarrow P_{x3}$ for different 
values of $n$, $\alpha$ and $\beta$ with some anisotropic effects as shown in 
the phase space portraits of Fig.~\ref{fig_4}. Here $x$ is a general index 
representing two indices $a$ and $b$. The plots show that the critical points 
lie within the value $1$. For $\alpha = 0.8$, $\beta = 1.275$, $n = 0.98$ the 
critical points are $P_{a1}(0,0.9845)$, $P_{a2}(0.968,0)$, $P_{a3}(0,0)$ and 
for $\alpha = 0.98$, $\beta = 1$, $n = 1$ the critical points are 
$P_{b1}(0,0.997)$, $P_{b2}(0.968,0)$, $P_{b3}(0,0)$. These non-unity value of 
critical points suggests the existence of anisotropy as predicted in the 
density parameter versus redshift plots in Fig.~\ref{fig3}. The fixed point
solutions for these two specific sets of model parameters are summarised in 
Table \ref{table5}.
%
\begin{center}
\begin{table}[h]
\caption{The fixed point solutions for the $f(Q)= -\beta Q^{n}$ model with 
two sets of constrained values of model parameters.}
\vspace{5pt}
\begin{tabular}{|c|c|c|c|c|c|c|c|c|}
\hline 
\rule[1ex]{0pt}{2.5ex}$\alpha$ &$\beta$& $n$&Fixed point &  $(x=\Omega_m,y=\Omega_r)$ &  $z=\Omega_\Lambda$ &  Eigenvalues  & $\omega_{eff}$   &  $q$  \\ 
\hline

\rule[1.25ex]{0pt}{2.5ex} & & &$P_{a1}$ &\Big($0,0.985$\Big) & $0.01$ & \Big($1,3.61$\Big) &$0.366$&$1.049$\\ 

\rule[1ex]{0pt}{2.5ex}$0.8$&$1.275$&$0.98$& $P_{a2}$ &\Big($0.968,0$\Big) & $0.026$ &($-\,1,2.61$) &$0.051$ &$0.576$\\ 

\rule[1ex]{0pt}{2.5ex}& & & $P_{a3} $ & \Big($0, 0$\Big) & $0.995$ & \Big($-\,3.61, -\,2.61$\Big) & $-\,0.835$&$-\,0.752$\\

\hline
\rule[1.25ex]{0pt}{2.5ex} & & &$P_{b1}$ &\Big($0,0.997$\Big) & $0.003$ & \Big($3.960, 1$\Big) &$0.33872$  &$1.008$\\ 

\rule[1ex]{0pt}{2.5ex}$0.98$&$1$&$1$& $P_{b2}$ &\Big($0.993,0$\Big) & $0.004$ &\Big($-\,1, 2.960$\Big)&$0.005$ &$0.507$\\ 

\rule[1ex]{0pt}{2.5ex}& & & $P_{b3} $ & \Big($0, 0$\Big) & $0.999$ & \Big($-\,3.960, -\,2.960$\Big)& $-\,0.982$ &$-\,0.972$\\ 
\hline
\end{tabular}
\label{table5}
\end{table}
\vspace{-27pt} 
\end{center}

\begin{figure}[!h]
\centerline{
  \includegraphics[scale = 0.4]{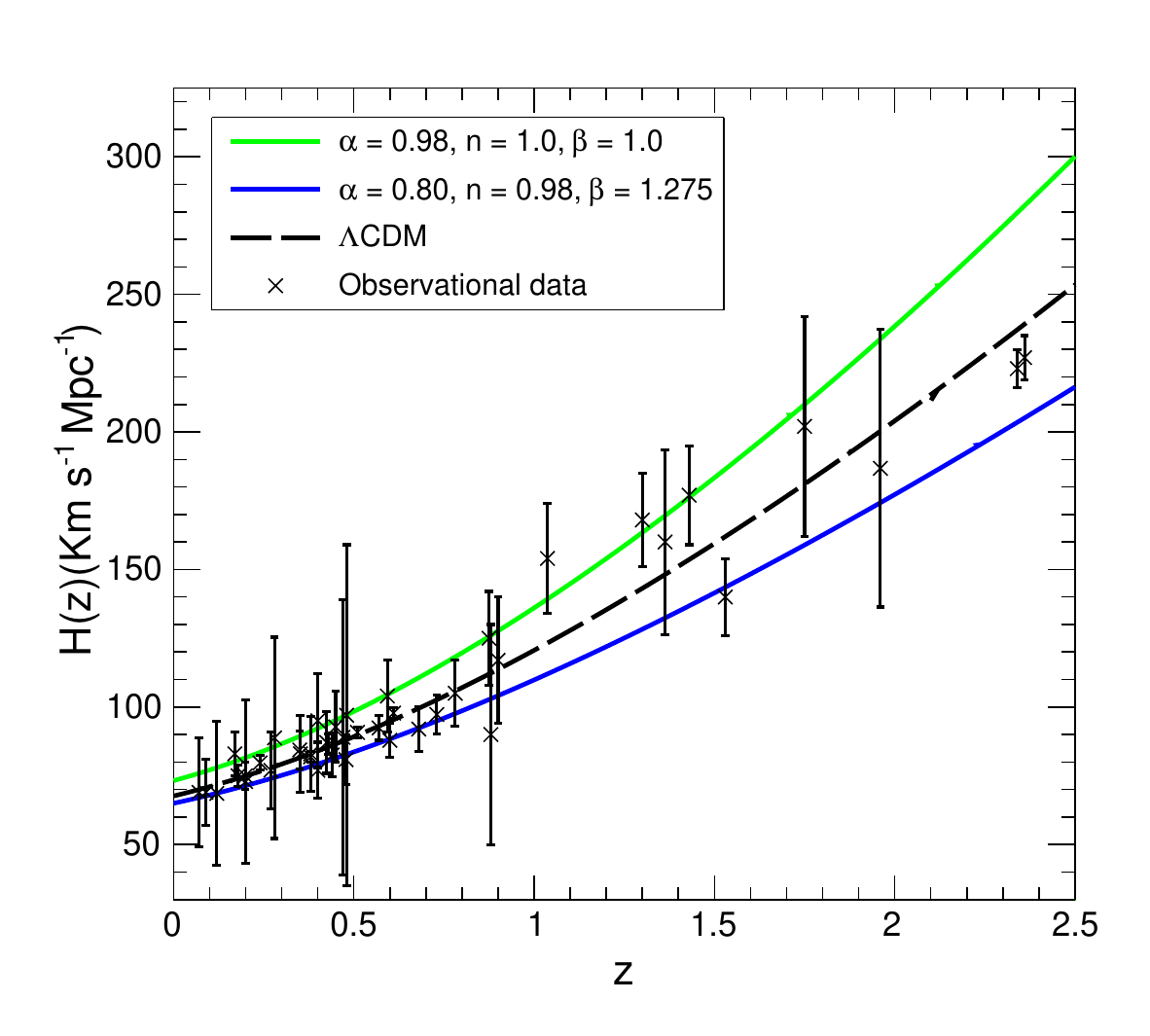}\hspace{0.25cm}
  \includegraphics[scale = 0.4]{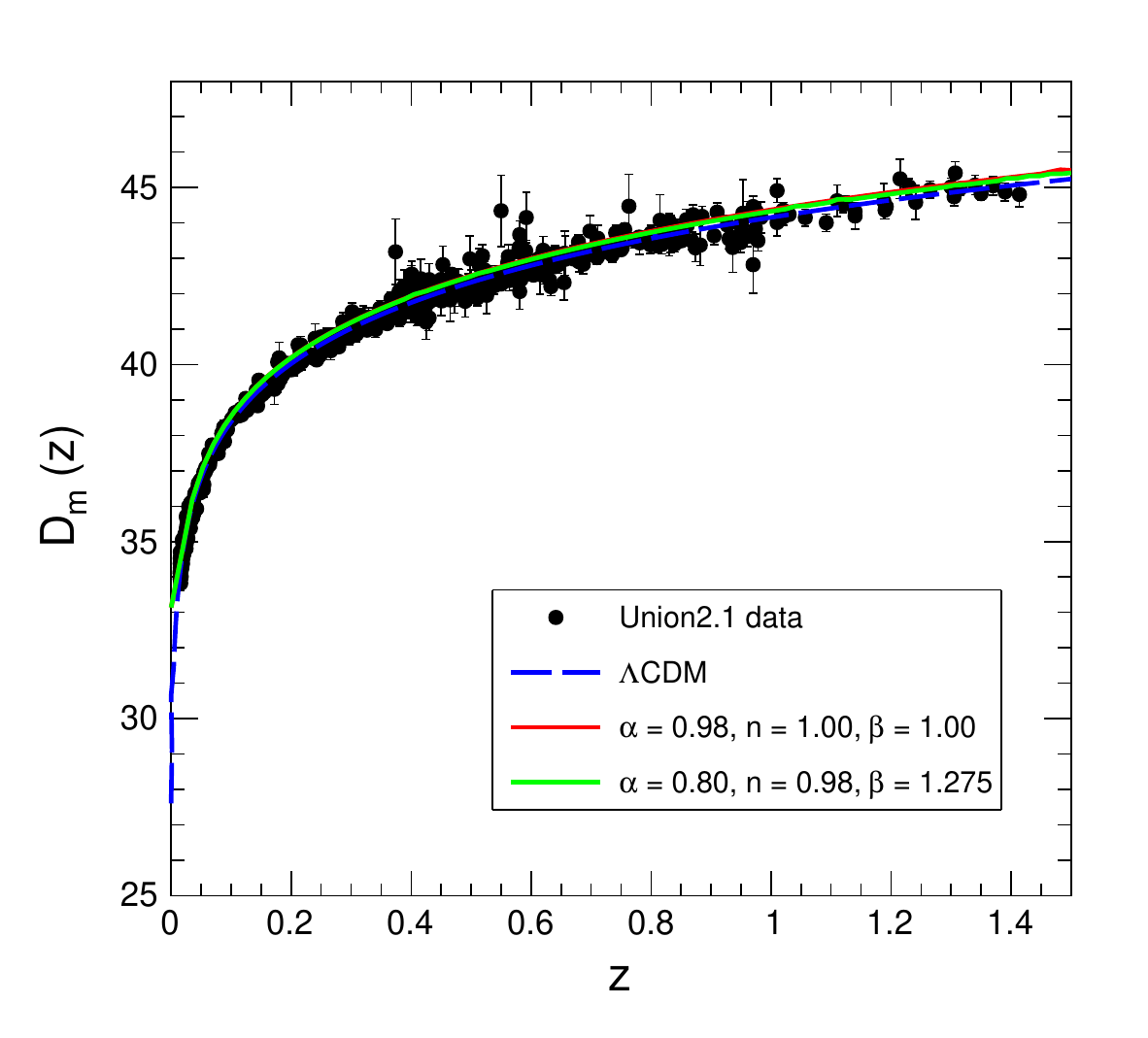}}
\vspace{-0.2cm}  
\caption{$H(z)$ versus $z$ and $D_m$ versus $z$ plots for the 
$f(Q) = \beta Q^{n}$ model with ($\alpha = 0.98, \beta = 1, n = 1$) and 
($\alpha = 0.8, \beta = 1.275, n = 0.98$) sets of model parameters in 
comparison with respective observational data.} 
\label{fig_5}
\end{figure}

We further test our constrained values of model parameters with the 
observational data of Hubble parameter ($H(z)$) from Table \ref{table6} and 
distance modulus ($D_m$) of Unioin 2.1 data \cite{Kowalski_2008,Amanullah_2010, Suzuki_2012} in Fig.~\ref{fig_5} and 
compare them also with the $\Lambda$CDM results. We use the expressions of 
Hubble parameter and distance modulus from the Ref.~\cite{Sarmah_2023}. The 
plots show that the constrained ranges of model parameters and anisotropy 
parameters agree with the observational data as well.
\begin{center}
\begin{table}[!h]
\caption{Available observational Hubble parameter ($H^{obs}(z)$) data set in 
the unit of km/s/Mpc obtained from different literature.}
\vspace{5pt}
\begin{tabular}{ccc|ccc}
\hline 
\rule[-1ex]{0pt}{2.5ex} \hspace{0.5cm} $z$ \hspace{0.5cm}  & \hspace{0.5cm} $ H^{obs}(z)$ \hspace{0.5cm} & \hspace{0.5cm} Reference \hspace{0.5cm} &
\hspace{0.5cm} $z$ \hspace{0.5cm}  & \hspace{0.5cm} $ H^{obs}(z)$ \hspace{0.5cm} & \hspace{0.5cm} Reference \hspace{0.5cm}\\ 
\hline
\rule[-1ex]{0pt}{2.5ex} 0.0708 & 69.0 $\pm$ 19.68 & \cite{Zhang_2014} & 
0.48 & 97.0 $\pm$ 62.0 & \cite{Ratsimbazafy_2017}\\
\rule[-1ex]{0pt}{2.5ex} 0.09 & 69.0 $\pm$ 12.0 & \cite{Simon_2005} &
0.51 & 90.8 $\pm$ 1.9 & \cite{Alam_2017}\\
\rule[-1ex]{0pt}{2.5ex} 0.12 & 68.6 $\pm$ 26.2 & \cite{Zhang_2014} &
0.57 & 92.4 $\pm$ 4.5 & \cite{Samushia_2013} \\
\rule[-1ex]{0pt}{2.5ex} 0.17 & 83.0 $\pm$ 8.0 & \cite{Simon_2005} &
0.593 & 104.0 $\pm$ 13.0 & \cite{Moresco_2012}\\
\rule[-1ex]{0pt}{2.5ex}0.179 & 75.0 $\pm$ 4.0 & \cite{Moresco_2012} &
0.60 & 87.9 $\pm$ 6.1 & \cite{Blake_2012}\\
\rule[-1ex]{0pt}{2.5ex} 0.199 & 75.0 $\pm$ 5.0 & \cite{Moresco_2012} &
0.61 & 97.8 $\pm$ 2.1 & \cite{Alam_2017}\\
\rule[-1ex]{0pt}{2.5ex} 0.20 & 72.9 $\pm$ 29.6 & \cite{Zhang_2014} &
0.68 & 92.0 $\pm$ 8.0 & \cite{Moresco_2012}\\
\rule[-1ex]{0pt}{2.5ex} 0.24 & 79.69 $\pm$ 2.65 & \cite{Gaztanaga_2009} &
0.73 & 97.3 $\pm$ 7.0 & \cite{Blake_2012}\\
\rule[-1ex]{0pt}{2.5ex} 0.27 & 77.0 $\pm$ 14.0 & \cite{Simon_2005} &
0.781 & 105.0 $\pm$ 12.0 & \cite{Moresco_2012}\\
\rule[-1ex]{0pt}{2.5ex} 0.28 & 88.8 $\pm$ 36.6 & \cite{Zhang_2014} &
0.875 & 125.0 $\pm$ 17.0 & \cite{Moresco_2012}\\
\rule[-1ex]{0pt}{2.5ex} 0.35 & 84.4 $\pm$ 7.0  & \cite{Xu_2013} &
0.88 & 90.0 $\pm$ 40.0 & \cite{Ratsimbazafy_2017}\\
\rule[-1ex]{0pt}{2.5ex} 0.352 & 83.0 $\pm$ 14.0 & \cite{Moresco_2012}&
0.90 & 117.0 $\pm$ 23.0 & \cite{Simon_2005}\\
\rule[-1ex]{0pt}{2.5ex} 0.38 & 81.9 $\pm$ 1.9 & \cite{Alam_2017} &
1.037 & 154.0 $\pm$ 20.0 & \cite{Moresco_2012}\\
\rule[-1ex]{0pt}{2.5ex} 0.3802 & 83.0 $\pm$ 13.5 & \cite{Moresco_2016} &
1.30 & 168.0 $\pm$ 17.0 & \cite{Simon_2005}\\
\rule[-1ex]{0pt}{2.5ex} 0.40 & 95.0 $\pm$ 17.0 & \cite{Simon_2005} &
1.363 & 160.0 $\pm$ 33.6 & \cite{Moresco_2015}\\
\rule[-1ex]{0pt}{2.5ex} 0.4004 & 77.0 $\pm$ 10.2 & \cite{Moresco_2016} &
1.43 & 177.0 $\pm$ 18.0 & \cite{Simon_2005}\\
\rule[-1ex]{0pt}{2.5ex} 0.4247 & 87.1 $\pm$ 11.2 & \cite{Moresco_2016} &
1.53 & 140.0 $\pm$ 14.0 & \cite{Simon_2005}\\
\rule[-1ex]{0pt}{2.5ex} 0.43 & 86.45 $\pm$ 3.68 & \cite{Gaztanaga_2009} &
1.75 & 202.0 $\pm$ 40.0 & \cite{Simon_2005}\\
\rule[-1ex]{0pt}{2.5ex} 0.44 & 82.6 $\pm$ 7.8 & \cite{Blake_2012} &
1.965 & 186.5 $\pm$ 50.4 & \cite{Moresco_2015}\\
\rule[-1ex]{0pt}{2.5ex} 0.4497 & 92.8 $\pm$ 12.9 & \cite{Moresco_2016} & 
2.34 & 223.0 $\pm$ 7.0 & \cite{Delubac_2015}\\
\rule[-1ex]{0pt}{2.5ex} 0.47 & 89.0 $\pm$ 50.0 & \cite{Ratsimbazafy_2017} &
2.36 & 227.0 $\pm$ 8.0 & \cite{Ribera_2014}\\
\rule[-1ex]{0pt}{2.5ex} 0.4783 & 80.9 $\pm$ 9.0 & \cite{Moresco_2016} &&&\\[2pt]
\hline
\end{tabular}
\label{table6}
\end{table}
\end{center}
\begin{center}
\begin{table}[!h]
\caption{The fixed points numerical solutions for $f(Q)= -\,\beta Q^n$ model 
for $n = 0.95$ with $\delta = ({\beta})^{0.05}$.}
\vspace{5pt}
\scalebox{0.90}{
\begin{tabular}{|c|c|c|c|c|c|c|c|c|}
\hline 
\rule[-1ex]{0pt}{2.5ex}$\alpha$& Fixed point&  $(x = \Omega_m,y= \Omega_r)$ &$z=\Omega_{\Lambda}$&  Eigenvalues  &  $\omega_{eff}$   &  $q$    \\ 
\hline

\rule[2ex]{0pt}{2.5ex} &$ P_{c11}$  &$\left(0,0 \right)$ &1& $\left[-4,-3\right]$&$0.05\, -0.98 \gamma$ &$0.58\, -1.46\gamma$\\ 

\rule[2ex]{0pt}{2.5ex} $\alpha = 1 $ &$ P_{c12}$  &$\left(0.99 \delta-0.05,0 \right)$ &$\left(1.05-{0.99}{\delta}\right)$&$\left[2.98 \left(\delta-1.39\right),5.96\left( \delta-0.55\right)\right]$&$0.94 \delta-{0.98}{\gamma}+0.005$ &$ 1.41 \delta-{1.46}{\gamma}+0.51$  \\ 

\rule[2ex]{0pt}{2.5ex}  &$ P_{c13}$  &$\left(0,0.74 \delta+0.2\right)$ &$\left(0.80-{0.74}{\delta} \right)$& $\left[2.98 \left( \delta-0.74\right),5.96 \left( \delta -0.40\right)\right]$& $0.94 \delta-{0.98}{\gamma}+0.31$ & $1.41 \delta-{1.46}{\gamma}+0.96$\\ 

\rule[-1ex]{0pt}{2.5ex}  & &&& &  & \\ 
\hline

\rule[2ex]{0pt}{2.5ex}  &$P_{c21}$&$\left( 0,0\right)$  & 0.96 & $[-3.1,-2.1]$&$0.25\, -{0.78}{\gamma}$&$0.87\, -{1.17}{\gamma}$\\

\rule[2ex]{0pt}{2.5ex}  $\alpha = 0.5$&$P_{c22}$&$\left(1.03 \delta-0.24,0\right)$&$\left(1.28-1.03 \delta\right)$&$\left[2.58 \left(\delta-1.47\right),5.15\left(\delta-0.68\right)\right]$&$0.82 \delta-{0.78}{\delta}+0.03$& $1.22\delta-{1.17}{\gamma}+0.54$\\ 

\rule[2ex]{0pt}{2.5ex} &$P_{c23}$&$\left( 0,0.77 \delta+0.08\right)$ &$\left(0.88-0.77 \delta\right)$& $\left[2.58 \left(1. \delta-0.72\right),5.15 \left(\delta-0.50\right)\right]$& $0.82\delta-{0.78}{\gamma}+0.33$&$1.22 \delta-{1.17}{\gamma}+0.99$\\ 
\rule[-1ex]{0pt}{2.5ex}  & &&& &  & \\ 
\hline
\rule[2ex]{0pt}{2.5ex}  &$P_{c31}$&$\left( 0,0\right)$  & 0.99 & $(-3.52,-2.52)$&$0.15\, -{0.89}{\gamma}$&$0.73\, -{1.33}{\gamma}$\\

\rule[2ex]{0pt}{2.5ex} $\alpha = 0.75$&$P_{c32}$&$\left(1.00 \delta-0.15, 0\right)$&$\left(1.15- \delta\right)$& $\left[2.75 \left( \delta-1.43\right),5.50 \left( \delta-0.61\right)\right]$&$0.87 \delta-{0.89}{\gamma}+0.02$&$1.31 \delta-{1.33}{\gamma}+0.52$\\ 

\rule[2ex]{0pt}{2.5ex} &$P_{c33}$&$\left(0, 0.75 \delta+0.14\right)$ &$\left(0.85- 0.75 \delta\right)$& $\left[2.75 \left( \delta-0.72\right),5.50\left( \delta-0.45\right)\right]$& $0.87 \delta-{0.89}{\gamma}+0.32$&$1.31 \delta-{1.3313}{\gamma}+0.98$\\ 
\rule[-1ex]{0pt}{2.5ex}  & &&& &  & \\ 
\hline

\rule[2ex]{0pt}{2.5ex}  &$P_{c41}$&$\left( 0,0\right)$  & 0.99& $[-4.52,-3.52]$&$-{1.05}{\gamma}-0.04$&$0.43\, -{1.58}{\gamma}$ \\

\rule[2ex]{0pt}{2.5ex}  $\alpha = 1.25$&$P_{c42}$&$\left( 0.99 \delta+0.04, 0\right)$&$\left(0.96 -0.99 \delta\right)$&$\left[3.24 \left( \delta-1.35\right),6.49 \left( \delta-0.50\right)\right]$&$1.03 \delta-{1.05}{\gamma}-0.003$ & $1.54 \delta-{1.58}{\gamma}+0.49$ \\ 

\rule[2ex]{0pt}{2.5ex} &$P_{c43}$&$\left( 0,0.75 \delta+0.25\right)$ &$\left( 0.76-0.75 \delta\right)$ &$\left[3.24 \left( \delta-0.75\right),6.49 \left( \delta-0.36\right)\right]$& $1.03 \delta-{1.05}{\gamma}+0.30$ & $1.54\delta-{1.58}{\gamma}+0.95$\\ 
\rule[-1ex]{0pt}{2.5ex} & & &&& &   \\ 
\hline

\end{tabular}}

\label{table3}
\end{table} 
\end{center}
\begin{center}
\begin{table}[!h]
\caption{The fixed points numerical solutions for $f(Q)= -\,\beta Q^n$ model for 
$n = 1.05$ with $\gamma =  ({\beta})^{-0.05}$}
\vspace{5pt}
\scalebox{0.90}{
\begin{tabular}{|c|c|c|c|c|c|c|c|c|}
\hline 
\rule[-1ex]{0pt}{2.5ex}$\alpha$& Fixed point&  $(x = \Omega_m,y= \Omega_r)$ &$z=\Omega_{\Lambda}$&  Eigenvalues  &  $\omega_{eff}$   &  $q$    \\ 
\hline

\rule[2ex]{0pt}{2.5ex} &$ P_{b11}$  &$\left(0,0 \right)$ &1& $\left[-4,-3\right]$&$-1.01 \gamma-0.05$ &$0.43\, -1.51 \gamma$\\ 

\rule[2ex]{0pt}{2.5ex} $\alpha = 1 $ &$ P_{b12}$  &$\left(0.05\, + 0.99 \gamma, 0 \right)$ &$\left(0.95-0.99\gamma\right)$&$\left[{5.96}\gamma-2.7,{2.98}{\gamma}-3.85\right]$&$0.03\gamma + 0.01$ & $0.06 \gamma+0.51$ \\ 

\rule[2ex]{0pt}{2.5ex}  &$ P_{b13}$  &$\left(0,0.3\, +{0.75}\gamma \right)$ &$\left(0.97-{0.75}{\gamma}\right) $& $\left[{5.96}{\gamma}-1.6,{2.98}{\gamma}-1.8\right]$&  ${0.03}{\gamma}+0.37$  & ${0.06}{\gamma}+1.06$\\ 

\rule[-1ex]{0pt}{2.5ex}  & &&& &  & \\ 
\hline

\rule[2ex]{0pt}{2.5ex}  &$P_{b21}$&$\left( 0,0\right)$  & 0.96 & $[-3.1,-2.1]$&$0.10\, -0.81 \gamma$&$0.64\, -1.21\gamma$\\

\rule[2ex]{0pt}{2.5ex}  $\alpha = 0.5$&$P_{b22}$&$\left({1.04}{\gamma}-0.13,0\right)$&$\left(1.09-{1.04}{\gamma}\right)$&$\left[{5.18}{\gamma}-2.73,{2.59}{\gamma}-3.41\right]$&${0.1}{\gamma}-0.02$&${0.05}{\gamma}+0.48$\\ 

\rule[2ex]{0pt}{2.5ex} &$P_{b23}$&$\left( 0,0.22\, +{0.78}{\gamma}\right)$ &$\left( 0.74\, -{0.78}{\gamma}\right)$& $\left[{5.18}{\gamma}-1.63,{2.59}{\gamma}-1.36\right]$ & ${0.1}{\gamma}+0.35$&${0.05}{\gamma} + 1.03$ \\ 
\rule[-1ex]{0pt}{2.5ex}  & &&& &  & \\ 
\hline
\rule[2ex]{0pt}{2.5ex}  &$P_{b31}$&$\left( 0,0\right)$  & 0.99 & $[-3.52,-2.52]$&$0.02\, -0.92 \gamma$&$0.54\, -1.37 \gamma$\\

\rule[2ex]{0pt}{2.5ex} $\alpha = 0.75$&$P_{b32}$&$\left({1.00}{\gamma}-0.03, 0\right)$&$\left(0.99- {1.00}{\gamma}\right)$&$\left[{5.51}{\gamma}-2.68,{2.76}{\gamma}-3.60\right]$&${0.04}{\gamma}-0.004$&${0.08}{\gamma}+0.49$\\ 

\rule[2ex]{0pt}{2.5ex} &$P_{b33}$&$\left(0, 0.26\, +{0.75}{\gamma}\right)$ &$\left(0.75\, -{0.75}{\gamma}\right)$& $\left[{5.51}{\gamma}-1.58,{2.76}{\gamma}-1.55\right]$& ${0.04}{\gamma}+0.37$&${0.08}{\gamma}+1.05$\\ 
\rule[-1ex]{0pt}{2.5ex}  & &&& &  & \\ 
\hline

\rule[2ex]{0pt}{2.5ex}  &$P_{b41}$&$\left( 0,0\right)$  & 0.99& $[-4.52,-3.52]$&$-1.09 \gamma-0.12$&$0.32\, -1.63 \gamma$ \\

\rule[2ex]{0pt}{2.5ex}  $\alpha = 1.25$&$P_{b42}$&$\left( 0.12\, +\gamma, 0\right)$&$\left( 0.88\, -\gamma\right)$&$\left[{6.50}{\gamma}-2.77,{3.25}{\gamma}-4.14\right]$&${0.05}{\gamma}+0.01$ & ${0.12}{\gamma}+0.52$ \\ 

\rule[2ex]{0pt}{2.5ex} &$P_{b43}$&$\left( 0,0.33\, +{0.75}{\gamma}\right)$ &$\left( 0.67\, -{0.75}{\gamma}\right)$& $\left[{6.50}{\gamma}-1.67,{3.25}{\gamma}-2.09\right]$ & ${0.05}{\gamma}+0.38$& ${0.12}{\gamma}+1.07$\\ 
\rule[-1ex]{0pt}{2.5ex} & & &&& &   \\ 
\hline

\end{tabular}}

\label{table4}
\end{table} 
\end{center}


\section{Dynamics of Anisotropy}\label{aniso}
To understand the dynamics of the density parameter of anisotropy, we consider
an approach in which rather than taking the assumption like in equation 
\eqref{H_directional}, a more generalised way the field equations 
\eqref{eom1}, \eqref{eom2} and \eqref{eom3} can be written without any extra 
parameter. For this, we re-write the field equations by considering the 
concept of average pressure $P = (p_x + 2 p_y)/3$ as follows:
\begin{equation}\label{fe_1}
\kappa \rho = \frac{f(Q)}{2} - f_Q Q,
\end{equation}
\begin{equation}\label{fe_2}
\kappa P = -\,\frac{f(Q)}{2} +2f_Q [3H^2 + 2\dot{H}]+ H \dot{Q}f_{QQ}.
\end{equation}
Moreover, we can express the field equations in terms of some dimensionless 
parameters. We define these parameters as
$$\eta= \frac{f}{6H^2f_Q},\; \Omega_m = \frac{\kappa \rho}{3H^2f_Q},\; 
\Omega_\sigma = \frac{\sigma^2}{3H^2},\; \text{and}\; 
m = \frac{Q f_{QQ}}{f_Q}.$$
With these parameters, equation \eqref{fe_1} takes the form:
\begin{equation}\label{om}
\Omega_m =  \eta + 2\Omega_\sigma - 2.
\end{equation}
Also, by using equation \eqref{fe_2} with the consideration of EoS 
$\omega = p/\rho$ we can write,
\begin{equation}
\frac{\dot{H}}{H^2}= \frac{\eta(1+\omega)-2\omega(1-\Omega_{\sigma})-2}{4mf_Q+\frac{2}{3}}.
\end{equation}
Hence, the effective EoS and the deceleration parameter take the forms 
respectively as
\begin{equation}\label{omef}
\omega_{eff}=-1-\frac{2}{3}\left[\frac{\eta(1+\omega)-2\omega(1-\Omega_{\sigma})-2}{4mf_Q+\frac{2}{3}}\right],
\end{equation}
\begin{equation}\label{dec}
q = -1-\frac{\dot{H}}{H^2}= -1-\left[\frac{\eta(1+\omega)-2\omega(1-\Omega_{\sigma})-2}{4mf_Q+\frac{2}{3}}\right].
\end{equation}
The first-order dynamical equations with respect to $N= \ln{a}$ for the 
considered set of dimensionless parameters can be written as
\begin{align}\label{dy_1}
\eta' & = 2\big[(1-\Omega_{\sigma})(1-m\eta)-\eta\big]\left[\frac{\eta(1+\omega)-2\omega(1-\Omega_{\sigma})-2}{4mf_Q+\frac{2}{3}}\right],\\[8pt]
\label{dy_2}
\Omega'_{\sigma} & = -2\Omega_{\sigma}\left[\frac{\eta(1+\omega)-2\omega(1-\Omega_{\sigma})-2}{4mf_Q+\frac{2}{3}}\right],\\[8pt]
\label{dy_3}
\Omega'_{m} & = 2\big[(1-\Omega_{\sigma})(1-m\eta)-\eta-\Omega_{\sigma}\big]\left[\frac{\eta(1+\omega)-2\omega(1-\Omega_{\sigma})-2}{4mf_Q+\frac{2}{3}}\right],
\end{align}
where the prime denotes the derivative with respect to $N$ as earlier.

Now, to avoid the complicacy we consider only our simple model 
$f(Q)= -(Q+2\Lambda)$ to do the dynamical system analysis with the present 
approach as an example. For this model, $f_{QQ}=0$ and hence $m=0$. Thus 
equation \eqref{dy_1}, \eqref{dy_2} and \eqref{dy_3} can be  re-written as
\begin{align}\label{dy1_mod1}
\eta' & = 3\big[(1-\Omega_{\sigma})-\eta\big]\big[{\eta(1+\omega)-2\omega(1-\Omega_{\sigma})-2}\big],\\[8pt]
\label{dy2_mod1}
\Omega'_{\sigma} & = -3\Omega_{\sigma}\big[{\eta(1+\omega)-2\omega(1-\Omega_{\sigma})-2}\big],\\[8pt]
\label{dy3_mod1}
\Omega'_{m} & = 3\big[(1-\Omega_{\sigma})-\eta-2\Omega_{\sigma}\big]\big[{\eta(1+\omega)-2\omega(1-\Omega_{\sigma})-2}\big].
\end{align}
Further, $\omega_{eff}$ and $q$ take the forms:
 \begin{align}\label{omef_mod1}
 \omega_{eff} & = -1 -\big[{\eta(1+\omega)-2\omega(1-\Omega_{\sigma})-2}\big],\\[8pt]
 \label{dec_mod2}
 q & = -1 - \frac{3}{2}\big[{\eta(1+\omega)-2\omega(1-\Omega_{\sigma})-2}\big].
 \end{align}
\begin{figure}[!htb]
\centerline{
  \includegraphics[scale = 0.4]{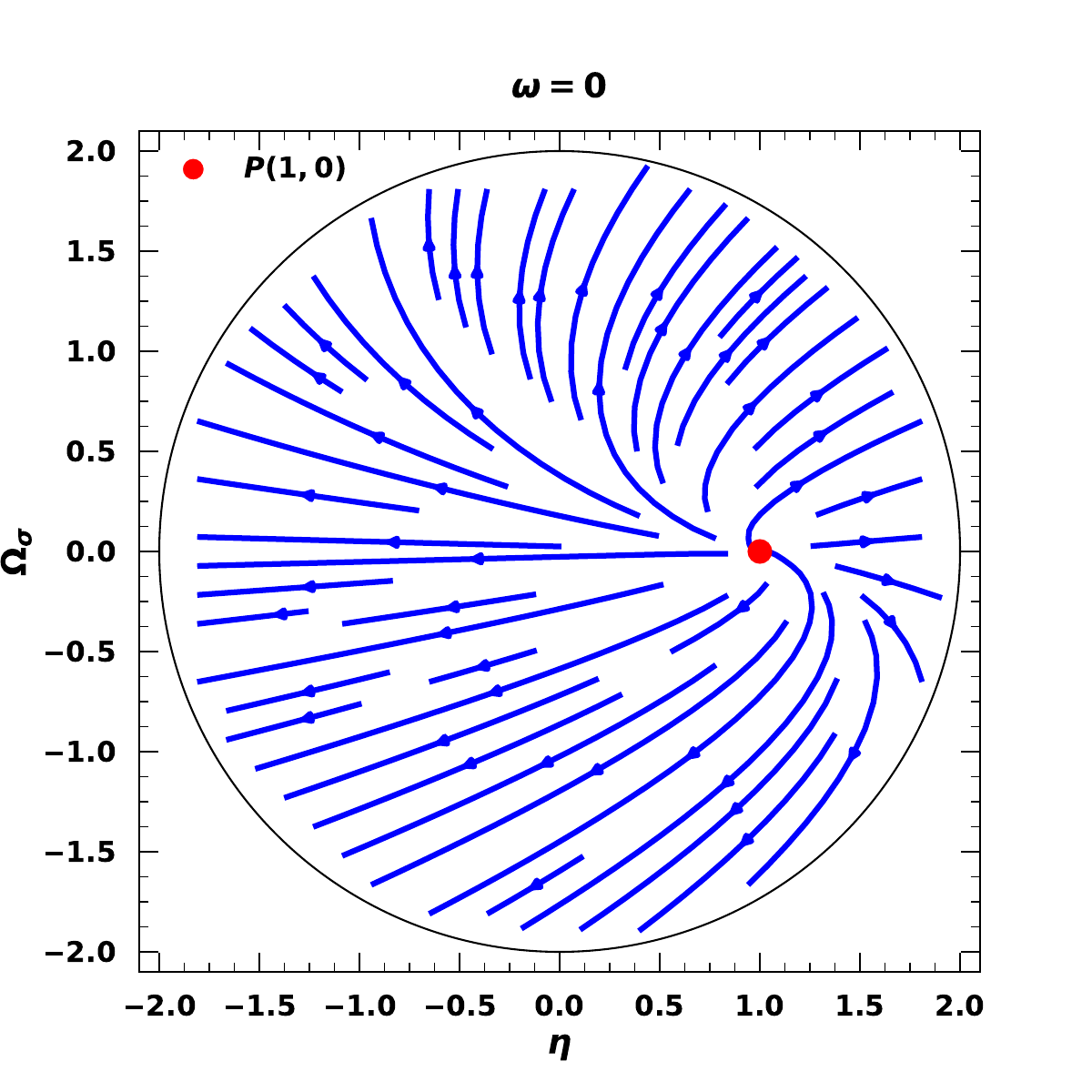}}
\vspace{-0.2cm}
\caption{Phase space portrait of $\eta$ vs $\Omega_{\sigma}$ for the
$f(Q) = -(Q+2\Lambda)$ model with $\omega =0$.}
\label{fig6}
\end{figure}
A simple phase space portrait for $\Omega_{\sigma}$ vs $\eta$ is shown in 
Fig.~\ref{fig6} for the $f(Q) = -(Q + 2\Lambda)$ model. The phase portrait 
shows a repeller or unstable node at $P(1,0)$ for $\omega = 0$. The dynamical 
system with $\Omega_{\sigma}$ and $\eta$ has the eigenvalue $(-3,3)$ which 
supports the existence of the unstable node in the dynamical system for the 
matter-dominated era.
\section{Conclusion}\label{con}
In this study, we have considered LRS Bianchi type I metric in the $f(Q)$ 
theory of gravity and treated it as a dynamical system.  Further, we have 
analysed the stability of the system by studying the critical point 
analysis for two different models: $f(Q) = -\,(Q+2\Lambda)$ and 
$f(Q) = -\,\beta Q^{n}$. 
During our study, we have found that in $f(Q) = -\,(Q+2\Lambda)$ model, the 
field equations mainly depend on $\alpha$, which is related to the directional 
anisotropy in the Hubble parameter. The model predicts the $\Lambda$CDM 
results for $\alpha = 1$, i.e.~for the isotropic case. Moreover, we have 
constrained the parameter $\alpha$ between $0.5$ to $1.25$ which agrees with 
the observational data as shown in Ref.~\cite{Sarmah_2023}. However, in phase 
space portrait, we have found that $\alpha > 1$ gives unphysical critical 
points. Thus it is suitable to set $\alpha$ within the range of $1$. For 
$\alpha < 1$ the model predicts radiation and matter-dominated phases along 
with dark energy-dominated phase with some anisotropic effects as the density 
parameters are not perfectly unity and these phases are heteroclinically 
connected. Further, the density parameters versus cosmological redshift plots 
suggest that the density parameters like $\Omega_m$ never reached unity due 
to the presence of anisotropy and also there is an elongated matter era as 
compared to standard $\Lambda$CDM prediction due to the presence of 
anisotropy.
    
For the model $f(Q) = -\beta Q^{n}$, we have found that the dynamical system 
depends not only on $\alpha$ but also on the model parameters $\beta$ and $n$. 
However, for $n = 1$, the system mainly depends on $\alpha$ only. For values
of $n$ other than $1$, the system depends on $\beta$ too. The density 
parameters versus cosmological redshift ($z$) plots help us to constrain our 
model parameters as $\alpha$ within the range of $(0.8 - 0.98)$, $\beta$ 
within the range of $(1 - 1.275)$ and $n$ in the range of $(0.98 - 1.00)$. For 
these values of model parameters, we have plotted the phase space portraits 
and found that the critical points lie within the range of unity. Like in the 
previous model, there are also three phases: radiation-dominated, 
matter-dominated, and dark energy-dominated eras with some anisotropic effect 
also observed and these phases are heteroclinically connected. Further, we 
have compared our Hubble and distance modulus plots for constrained values of 
model parameters with the observational data and standard $\Lambda$CDM results. We have found that these plots are in good agreement with observational data. 
In this model too we have noticed an elongated matter-dominated era due to the 
presence of anisotropy as compared to $\Lambda$CDM prediction.
  
From this study, we have found that for both models the anisotropic parameter 
$\alpha < 1$ gives only the physically viable critical points and $\alpha = 1$ 
gives isotropic Universe with $\Lambda$CDM results by the 
$f(Q) = (Q + 2\Lambda)$ model, and with $n = 1, \beta = 1$ along with 
$\alpha = 1$ gives the similar results by the $f(Q) = - \beta Q^{n}$ model too.
Both models suggest some role of anisotropic effect in the cosmic evolution of 
the Universe and the contribution to density parameters in the different 
phases of the Universe. 
  
In addition to the above-mentioned studies, we have also tried to study the 
dynamics of anisotropy in Section \ref{aniso} in which we have considered the 
anisotropy as a dynamical variable with the consideration of the concept of 
average pressure $P = (P_x + 2P_y)/3$. We have drawn a phase space portrait 
for $\Omega_{\sigma}$ vs $\eta$ with $\omega = 0$ for the first model in 
Fig.~\ref{fig6}, where $\eta= f/6H^2f_Q$. The portrait shows an unstable node 
at $P(1,0)$ with the eigenvalue ($-\,3,3$). We have restricted ourselves to 
studying the dynamics of the anisotropy for radiation and dark energy phase as 
the equations are more complicated in this approach. We may cover this study 
in our later work with some detailed analysis.

{In simple words, our analysis of the two models mentioned above using 
the LRS-BI Universe as a dynamical system in $f(Q)$ gravity allows us to 
summarize our findings as that both models point to the influence of 
anisotropy on the length of distinct phases of the Universe that are 
heteroclinically connected and therefore, have an impact on the Universe's 
cosmic evolution.}

The dynamical system analysis of a cosmological system provides information 
about the various evolutionary phases of the Universe and help us to 
understand the stability of the system through the study of the nature of the 
critical points and their eigenvalues. In our work, we have found that there 
is a heteroclinic sequence followed by the critical points together with some 
anisotropic effects while evolving the Universe through radiation-dominated, 
matter-dominated, and dark energy-dominated phases.

\section*{Acknowledgements} UDG is thankful to the Inter-University Centre for
Astronomy and Astrophysics (IUCAA), Pune, India for the Visiting
Associateship of the institute.

\bibliographystyle{apsrev}

\newpage

\end{document}